\documentclass[twocolumn,trackchanges]{aastex701}
\shorttitle{Constraints on massive CE phase}
\shortauthors{Z. Li et al.}

\begin{document}

\title{A Path to Constraints on Common Envelope Ejection in Massive Binaries: Full Evolutionary Reconstruction of Three Black Hole X-ray Binaries}

\correspondingauthor{Z. Li; H. Ge; X. Chen}
%\email{lizw@ynao.ac.cn; cxf@ynao.ac.cn}

\author[0000-0002-1421-4427]{Zhenwei Li}
\affiliation{International Centre of Supernovae (ICESUN), Yunnan  Key Laboratory of Supernova Research, Yunnan Observatories, Chinese Academy of Sciences (CAS), Kunming 650216, People's Republic of China}
\affiliation{University of Chinese Academy of Sciences, Beijing 100049, People's Republic of China}
\email[show]{lizw@ynao.ac.cn}

\author[0000-0003-0821-4583]{Dandan Wei}
\affiliation{Institute of Science and Technology Austria: Klosterneuburg, Lower Austria, AT}
\email{dandan.wei@hotmail.com}

\author[0000-0001-6808-638X]{Shi Jia}
\affiliation{International Centre of Supernovae (ICESUN), Yunnan  Key Laboratory of Supernova Research, Yunnan Observatories, Chinese Academy of Sciences (CAS), Kunming 650216, People's Republic of China}
\affiliation{University of Chinese Academy of Sciences, Beijing 100049, People's Republic of China}
\email{jiashi@ynao.ac.cn}

\author[0009-0006-9211-2860]{Hailiang Chen}
\affiliation{International Centre of Supernovae (ICESUN), Yunnan  Key Laboratory of Supernova Research, Yunnan Observatories, Chinese Academy of Sciences (CAS), Kunming 650216, People's Republic of China}
\affiliation{University of Chinese Academy of Sciences, Beijing 100049, People's Republic of China}
\email{chenhl@ynao.ac.cn}

\author[0000-0002-1421-4427]{Hongwei Ge}
\affiliation{International Centre of Supernovae (ICESUN), Yunnan  Key Laboratory of Supernova Research, Yunnan Observatories, Chinese Academy of Sciences (CAS), Kunming 650216, People's Republic of China}
\affiliation{University of Chinese Academy of Sciences, Beijing 100049, People's Republic of China}
\email[show]{gehw@ynao.ac.cn}

\author[0000-0001-7420-9606]{Zhuo Chen}
\affiliation{Institute for Advanced Study, Tsinghua University, Beijing 100084, China}
\email{chenzhuo_astro@tsinghua.edu.cn}

\author[0009-0008-3363-265X]{Yangyang Zhang}
\affiliation{Zhoukou Normal University, East Wenchang Street, Chuanhui District, Zhoukou, 466001, People's Republic of China}
\email{zhangyy@zknu.edu.cn}

\author[0000-0001-5284-8001]{Xuefei Chen}
\affiliation{International Centre of Supernovae (ICESUN), Yunnan  Key Laboratory of Supernova Research, Yunnan Observatories, Chinese Academy of Sciences (CAS), Kunming 650216, People's Republic of China}
\affiliation{University of Chinese Academy of Sciences, Beijing 100049, People's Republic of China}
\email[show]{cxf@ynao.ac.cn}

\author[0000-0001-9204-7778]{Zhanwen Han}
\affiliation{International Centre of Supernovae (ICESUN), Yunnan  Key Laboratory of Supernova Research, Yunnan Observatories, Chinese Academy of Sciences (CAS), Kunming 650216, People's Republic of China}
\affiliation{University of Chinese Academy of Sciences, Beijing 100049, People's Republic of China}
\email{zhanwenhan@ynao.ac.cn}
%\nocollaboration


\begin{abstract}
  The massive binary common envelope (CE) phase plays a pivotal role in the formation of close black hole/neutron star (BH/NS) binaries, yet significant uncertainties remain in our understanding of this process. In this study, we aim to constrain the massive binary CE phase by systematically reconstructing three observed BH X-ray binaries (BHXBs): GRO J1655-40, SAX J1819.3-2525, and 4U 1543-47. Through comprehensive binary evolution simulations and parametric supernova modeling, we establish lower limits for the CE efficiency parameters under different energy considerations within the standard energy formalism. Specifically, we derive minimum values for three cases: $\alpha_{\rm 0.5U}$ and $\alpha_{\rm U}$ representing CE efficiencies with half and all of the internal energy contributing to the envelope ejection, respectively, and $\alpha_{\rm H}$ accounting for the envelope's enthalpy. Our analysis reveals that the self-consistent formation of these three BHXBs requires CE efficiency parameters satisfying: $\alpha_{\rm 0.5U}\gtrsim 6.7$, $\alpha_{\rm U}\gtrsim 4.2$ and $\alpha_{\rm H}\gtrsim 1.7$. Notably, we find no viable solutions with CE efficiency values below unity, even when considering the most extreme scenarios in which the envelope binding energy is significantly reduced through enthalpy inclusion. {Our results strongly imply that either additional energy sources are required, or the formalism itself must be revised.} Furthermore, we quantitatively assess the impact of BH natal kicks on our results. A key finding is that 4U 1543-47’s formation requires substantial natal kicks ($\gtrsim 50 \;\rm km/s$), as lower kick velocities are incompatible with isolated binary evolution.
\end{abstract}

%\emph{Unified Astronomy Thesaurus concepts:} Neutron star (1108); Binary pulsars (153); Millisecond pulsars (1062)



\section{Introduction}
\label{sec:1}

Binary star systems are ubiquitous in the universe, with observational studies suggesting that a significant fraction of stellar systems exist in binary or multiple configurations \citep{Sana2012,Moe2017,Guoy2022,Chenx2024}. The evolution of binary systems plays a pivotal role in understanding various astrophysical phenomena, including the formation of compact object binaries, type Ia supernovae (SN) progenitors, and X-ray binary systems \citep{DeMarco2017,Hanz2020,Liuz2023,Tauris2023,Belloni2023,Chenx2024}. Despite remarkable theoretical and observational advances in recent decades, fundamental challenges remain in our understanding of binary evolution processes. Among these, common envelope (CE) evolution stands out as one of the most critical yet poorly understood phases, serving as a key mechanism for producing cataclysmic variables, ultra-compact X-ray binaries, and binary gravitational wave sources (see reviews by \citealt{Ivanova2013,Ivanova2020,Ropke2023}).

Although significant effort has been devoted to modeling CE evolution through numerical simulations spanning 1D to 3D approaches \citep[e.g.,][]{Soker2016,Ohlmann2016,Ivanova2016,Kruckow2016,Iaconi2017,Shiber2018,Chamandy2018,Fragos2019,Shiber2019,Law-Smith2020,Grichener2021,Moreno2022,Lau2022,Bronner2024,Gagnier2025,Vetter2025}, the astrophysical community has largely adopted a phenomenological description based on energy conservation as a practical framework for modeling the CE ejection \citep{Paczynski1976,Webbink1984,Livio1988}. In this energy formalism\footnote{{In the remainder of this paper, the CE energy formalism refers to the standard version used in binary population synthesis studies. It therefore does not consider additional processes or energy sources that may operate during the CE phase, including but not limited to jets (e.g., \citealt{Soker2004,Soker2015,Soker2016,Shiber2019}), magnetic fields (e.g., \citealt{Ohlmann2016,Garcia2020}) and nuclear energy \citep{Podsiadlowski2010}.}}, the system's orbital energy release is assumed to overcome the envelope's binding energy. This approach offers significant advantages: its simplicity and computational efficiency make it particularly suitable for population synthesis studies (see recent reviews by \citealt{Hanz2020,Chenx2024}).
%Despite a mount of nummerical simulations on CE evolution, including from one-dimensional to three-dimensional simulations \citep[e.g.,][]{Ohlmann2016,Ivanova2016,Kruckow2016,Chamandy2018,Fragos2019,Law-Smith2020,Lau2022,Bronner2024,Gagnier2025,Vetter2025}, no definitive conclusion has been made on this crucial process. It is widely accepted to adopt the phenomenological description based on energy conservation to simplify the CE ejection processes \citep{Paczynski1976,Webbink1984,Livio1988}, i.e. the so-called energy mechanism. In the energy mechanism, the release of orbital energy is used to eject the binding energy of the envelope. The energy mechanism is simple and has strong operability, thus being fit for rapid population synthesis studies (see \citealt{Belloni2023,Chenx2024} for recent reviews). 

The CE ejection efficiency $\alpha_{\rm CE}$, representing the fraction of orbital energy actually used to eject the envelope, remains one of the most crucial yet poorly constrained parameters in CE evolution. As a free parameter in theoretical models, $\alpha_{\rm CE}$ proves difficult to determine from numerical simulations alone. Analysis of post-CE systems, especially close white dwarf and hot subdwarf binaries, currently provides the most reliable constraints on the CE efficiency \citep[e.g.,][]{Zorotovic2010,Rebassa2012,DeMarco2011,Davis2012,Iaconi2019,Zorotovic2022,Hernandez2022,Scherbak2023,Geh2022,Belloni2024a,Belloni2024b,Geh2024,Yamaguchi2024,Zhangy2024,Torres2025}. These analyses reveal two consistent trends: (1) the majority of post-CE WD binaries require only modest $\alpha_{\rm CE}$ values ($<1$), (2) the efficiency appears to vary systematically with progenitor system properties, showing strong dependencies on both initial masses and orbital periods \citep{DeMarco2011,Davis2012,Zhangy2024,Geh2022,Geh2024}.

Post-CE systems containing white dwarfs or hot subdwarfs primarily originate from binaries with intermediate- and low-mass donors ($\lesssim 8\,M_\odot$). In contrast, CE evolution involving massive stars ($\gtrsim 8\,M_\odot$) presents significantly greater complexity due to both theoretical and observational challenges. Theoretically, uncertainties persist regarding the structure and evolution of massive stars, including (but not limited to) their stellar winds, SN explosion mechanisms, and compact remnant natal kicks \citep{Vink2012,Vink2022,Marchant2024,Weid2024}. Observationally, post-CE systems descended from massive progenitors remain exceptionally rare \citep{Gotberg2023,Yangz2025}. These combined factors have led to particularly poor constraints on CE evolution in massive binary systems.

%Refer to Belloni
Current understanding suggests that massive stellar envelopes are significantly more difficult to eject than their low-mass counterparts \citep{Podsiadlowski2003,Pfahl2003,Justham2006,Kiel2006,Yungelson2006,Podsiadlowski2010,Zuoz2014,Wiktorowicz2014,Lix2015,Shaoy2015,Wangc2016a,Wangc2016b,Fragos2019,Wilson2022}. This challenge primarily stems from the substantial radiative envelopes surrounding massive stellar cores, which dramatically increase the total binding energy \citep{Wangc2016a,Wangc2016b,Xux2010,Hirai2022,Wilson2022,Picker2024}. To account for this difficulty, binary population synthesis studies of massive systems often employ (equivalent) CE efficiency\footnote{The binding energy parameter $\lambda$ is commonly introduced, with the product $\alpha_{\rm CE}\lambda$ representing the combined effects of envelope binding energy and ejection efficiency \citep{deKool1987}. Some studies adopted high $\lambda$ values while maintaining modest $\alpha_{\rm CE}$.} values as high as $\alpha_{\rm CE} \sim 3-10$ \citep{Dominik2012,Briel2023,Grichener2023,Tanaka2023,Dengz2024a,Niey2025}. Notably, the adopted CE efficiency dramatically impacts predicted merger rates of double compact objects, introducing uncertainties spanning 2-3 orders of magnitude \citep{Dominik2012,Kruckow2018,Grichener2023,Tanaka2023}. It becomes fundamentally important to constrain CE efficiency parameters in massive binary systems. 

%Recently, Fragos+2019 performed the complete evolution of the CE ejection processes for a $12M_\odot$ red supergiant and a NS with 1d hydrodynamic simulations. They found that the CE efficiency for massive binaries may be as high as $5$. 

In this study, we establish constraints on the CE efficiency for massive binaries by analyzing three observed black hole intermediate-mass X-ray binaries (BH IMXBs) - systems believed to be products of CE evolution (see Section~\ref{sec:2} for detailed descriptions). Our approach combines detailed binary evolution simulations with a parametric SN model, enabling us to reconstruct the complete evolutionary history of these selected BH IMXBs in a self-consistent framework. The derived lower limits on CE efficiency values will provide crucial inputs for future population synthesis studies of massive binary systems.

The remainder of this paper is structured as follows. Section \ref{sec:2} details our simulation methods and key input parameters. In Section \ref{sec:3}, we present the simulated parameter grids for the three BH X-ray binaries (BHXBs), derive constraints on CE efficiency values, and analyze the impact of BH natal kicks. Finally, Sections \ref{sec:4} and \ref{sec:5} discuss our findings and present conclusions, respectively.

\begin{deluxetable*}{ccccccc}
  \tablecaption{The binary parameters of the three BHXBs\label{tab:1}}
\tablecolumns{1}
\tablenum{1}
\tablewidth{0pt}
\tablehead{
\colhead{Name} &
\colhead{$P_{\rm orb}(\rm h)$} &
\colhead{$M_{\rm BH}(M_\odot)$} &
\colhead{$M_{2}(M_\odot)$} &
\colhead{$T_{\rm eff}(\rm K)$} &
\colhead{$a_{\rm BH}$} &
\colhead{Refs}
}
\startdata
GRO J1655-40$^{*}$ & $62.920$ &  $6.3\pm 0.5$ & $2.4\pm0.4$ & $5715-5990$ & $0.7\pm0.05$ & (1)\\
SAX J1819.3-2525 & 67.6152 &  $6.4\pm0.6$ & $2.9\pm0.4$ & $12261-12461$ & - & (2)\\
4U 1543-47 & $26.79377$ &  $9.4\pm2.0$ & $2.63\pm0.24$ & $9000\pm500$ & $0.67^{+0.15}_{-0.08}$& (3)\\
\enddata
\tablecomments{$^*$ \citet{Beer2002} suggested another set of constraints with $M_{\rm BH}=5.4\pm0.4\,M_\odot$ and $M_{\rm 2}=1.45\pm0.35\,M_\odot$ for GRO J1655-40 (see also \citealt{Motta2014}). (1) \citet{Orosz1997,Greene2001,Shahbaz2003,Shafee2006}; (2) \citet{Orosz2001,MacDonald2014}; (3) \citet{Orosz1998,Orosz2003,Tetarenko2016,Dongy2020}.
}
\end{deluxetable*}

\section{Methods}
\label{sec:2}
\subsection{Observed parameters of the chosen BHXBs}
\label{sec:2.1}

This study investigates the CE phase in massive binaries through analysis of three dynamically confirmed BH IMXBs with well-constrained system parameters (Table~\ref{tab:1}), i.e., GRO J1655-40 (hereafter GRO J1655), SAX J1819.3-2525 (hereafter SAX J1819) and 4U 1543-47 (hereafter 4U 1543). These systems were selected based on the following criteria: (1) Galactic disk localization \citep{Corral2016,Tetarenko2016,Fortin2024}, suggesting isolated binary evolution origins; (2) Extreme initial mass ratios of the primordial binaries indicating probable CE evolution (see Section \ref{sec:2.2} for details). Our focus on BH IMXBs rather than BH low-mass XBs (BH-LMXBs; \citealt{Liuq2007,Corral2016}) is motivated by two key factors: (1) {the magnetic braking processes play a vital role in the formation of BH-LMXBs and the uncertain magnetic braking processes introduce significant errors in post-CE orbital separation estimates} (Section~\ref{sec:2.3}; \citealt{Dengz2024b}); (2) Substantial mass loss in low-mass companions \citep{Chenw2006,Justham2006,Lix2008} and associated progenitor mass uncertainties would dramatically increase computational requirements.

%In this work, we focus on three dynamical confirmed BH-IMXBs with well constraint binary parameters to study the CE phase of massive binaries. The observed parameters of these three BHXBs are shown in Table~\ref{tab:1}. The reason to choose these three BHXBs are introduced as follows\footnote{Most BHXBs are found with low-mass companions, i.e., BH-LMXBs \citep{Liuq2007,Corral2016}. We do not chose BH-LMXBs as the research objective because of two reasons. First, the large uncertainties of magnetic braking processes during the formation of BH-LMXBs make the estimation of post-CE binary separations being inaccurate (see Section 2.3 for more details; \citealt{Dengz2024b}). Second, the low-mass non-degenerate star may have experienced significant mass loss \citep{Justham2006,Lix2008}, and the large uncertainty of its progenitor mass would largely enlarge the calculation amount.}. (1) They are likely located in the Galaxy disc \citep{Corral2016}, which makes them more likely being formed from isolated binary evolution. (2) They are more likely formed with experiencing the CE ejection process due to the large initial mass ratios of the progenitor binaries (see Section X for more details).  

\subsection{Formation scenario of BHXBs}
\label{sec:2.2}

\begin{figure*}
    \centering
    \includegraphics[width=0.8\textwidth]{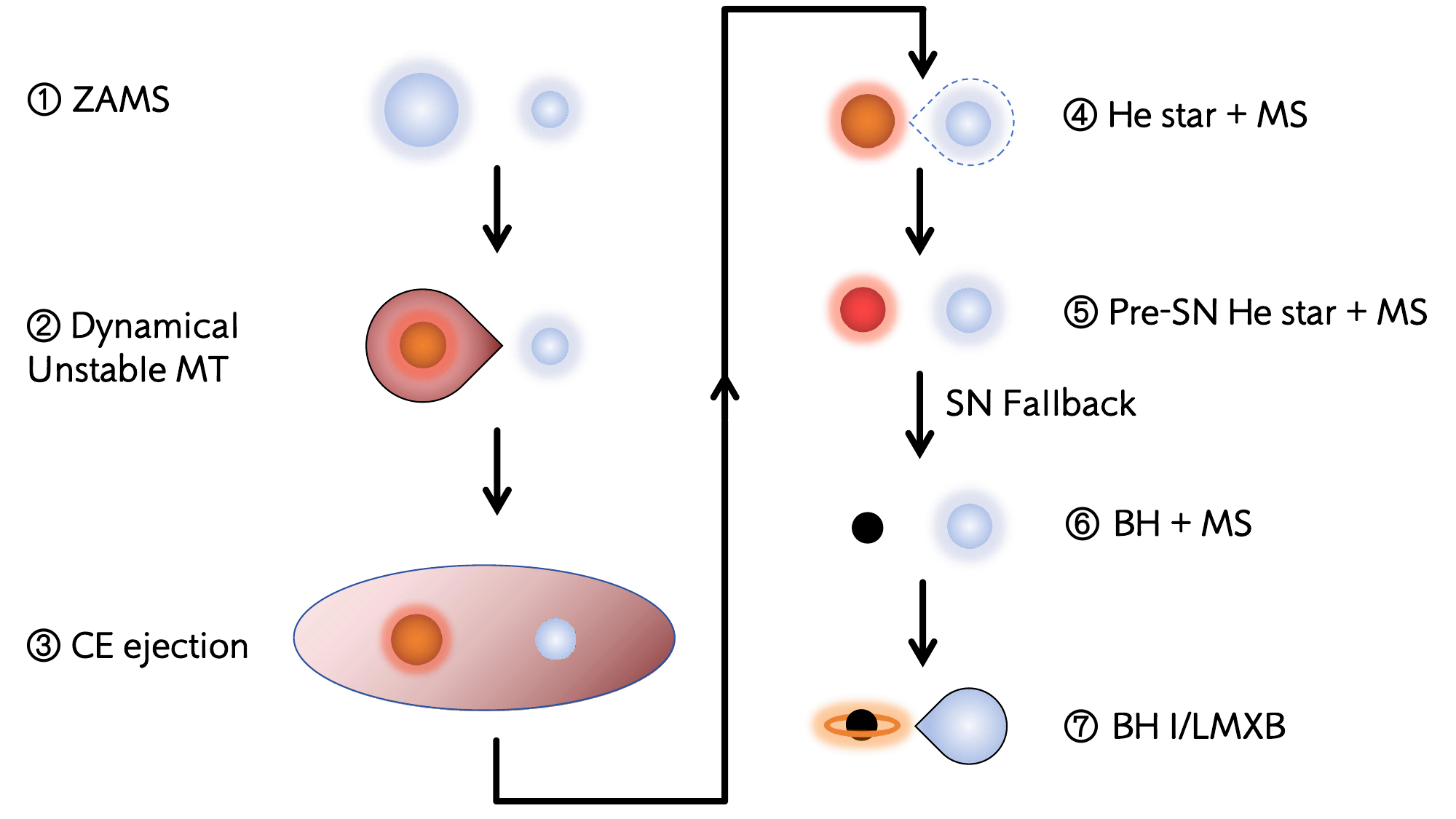}
    \caption{The typical evolutionary scenario for BH I/LMXBs. Abbreviations are as follows: ZAMS--zero-age main sequence, MT--mass transfer, CE--common envelope, He--helium, MS--main sequence, SN--supernova, BH--black hole, I/LMXB--intermediate-/low-mass X-ray binary. {At stage 4, the dashed contour denotes the Roche lobe of the secondary star. Note that the successful ejection of the CE requires the secondary’s radius to be within its Roche lobe.}}
    \label{fig:1}
\end{figure*}

Figure~\ref{fig:1} presents the typical evolutionary sequence for BHXBs formation through isolated binary evolution, beginning with a primordial system containing a massive primary ($M_{\rm 1}\gtrsim 32\,M_\odot$, based on the parametric SN models; Section~\ref{sec:2.4}) and a lower-mass secondary ($M_{2}\lesssim 6.2\,M_\odot$, from the binary simulations below in Section~\ref{sec:3.1}; \textbf{Stage 1}). The primary evolves off the MS after forming a helium (He) core, initiating Roche lobe overflow with a mass ratio $ \gtrsim 5$ (\textbf{Stage 2}) that leads to dynamically unstable mass transfer \citep{Geh2020a,Geh2020b,Geh2025} and a subsequent CE phase (\textbf{Stage 3}). Successful CE ejection - an essential requirement for BHXB formation - leaves a stripped He star and a nearly unevolved MS companion (\textbf{Stage 4}). The He star then loses significant mass through stellar winds (\textbf{Stage 5}) before undergoing core collapse to form a BH via SN fallback (\textbf{Stage 6}). Finally, as the secondary evolves and fills its Roche lobe, accretion onto the BH produces X-ray emission, completing the formation of an observable BHXB system (\textbf{Stage 7}).

%In Figure~\ref{fig:1}, we present the typical evolutionary scenario of the three selected BHXBs from an isolated binary system. The initial primodial binary contains a primary star with mass of $32-40M_\odot$ and a secondary star with mass of $2-7M_\odot$, where the primary mass are obtained based on the supernova explosion simulations in Schneider+2021 and the secondary mass are from our binary simulation results (see Section xx for more details). The massive one evolves first and fills its Roche lobe at a certain evolutionary stage (beyond the MS stage where the helium core has been born). The mass ratio at this moment is still large enough (typically larger than $4$ based on our simulations; see section xx) that the binary would definately enters into the CE phase. Assumed that the CE can be successfully ejected (otherwise the BHXBs will not be produced), the binary leaves a He star and an almost unevolved MS star. The He star then evolves and loses a significant part of masses due to the strong wind, finally it collapses to an black hole. The secondary star now begins to expand and fills its Roche lobe. The mass accretion of BH would emit the X-ray, and the BHXBs are produced eventually. 

\subsection{The technical route}
\label{sec:2.3}

The formation of BHXBs\footnote{Throughout the remainder of this paper, the term BHXB refers specifically to BH IMXB.} requires the progenitor binary to undergo a CE phase, which is generally described by the classical energy formalism \citep{Webbink1984,Livio1988}. This mechanism relates the envelope's binding energy to the change in orbital energy through:
\begin{eqnarray}
  E_{\rm bind} = \alpha_{\rm CE}\left(-\frac{GM_{\rm 1,s3}M_{\rm 2}}{2a_{\rm pre-CE}}+\frac{GM_{\rm 1c,s3}M_{\rm 2}}{2a_{\rm post-CE}}\right),
  \label{eq:1}
\end{eqnarray}
where $E_{\rm bind}$ represents the envelope binding energy at the onset of CE phase, $M_{\rm 1,s3},\;M_{\rm 1c,s3}$ are for the masses of the primary and primary's core at Stage 3, respectively, $a_{\rm pre-CE},\; a_{\rm post-CE}$ are the orbital separation before and after the CE ejection phase. To constrain the CE efficiency $\alpha_{\rm CE}$, we must determine the binary parameters at two critical evolutionary stages: (1) pre-CE phase when the primary fills its Roche lobe (Stage 3); (2) post-CE phase consisting of a newly formed He star and an (nearly) unevolved MS companion (Stage 4). Our methodology integrates both detailed single star evolutionary calculations and comprehensive binary star simulations to consistently model the system's evolution across these key phases.

\textbf{Step 1 - Determining Pre-CE Binary Parameters ($M_{\rm 1,s3}$, $M_{\rm 1c,s3}$, $a_{\rm pre-CE}$, $E_{\rm bind}$):} We first conduct single star simulations for the BH progenitor (see Section 2.4). When the star develops a He core and initiates Roche lobe overflow, we calculate the pre-CE orbital separation ($a_{\rm pre-CE}$) via \citep{Eggleton1983}
\begin{eqnarray}
  a_{\rm pre-CE} = R_{\rm 1,L}\frac{0.6q_{\rm s3}^{2/3}+\ln(1+q_{\rm s3}^{1/3})}{0.49q_{\rm s3}^{2/3}},
  \label{eq:2}
\end{eqnarray}
where $R_{\rm 1,L}\approx R_{\rm 1,s3}$ and $R_{\rm 1,s3}$ is the primary's radius at Stage 3, $q_{\rm s3} = M_{\rm 1,s3}/M_{\rm 2}$ is the mass ratio and $M_{\rm 2}$ is the companion mass determined from binary evolution simulations (described below). 

We employ three distinct approaches to compute the envelope binding energy ($E_{\rm bind}$) of the donor star at Stage 3. For Cases 1 and 2, we calculate $E_{\rm bind}$ using the standard energy integral: 
\begin{eqnarray}
  E_{\rm bind} = -\int^{\rm surface}_{\rm core}\left(-\frac{Gm}{r}+\alpha_{\rm th}U\right){\rm d} m,
  \label{eq:3}
\end{eqnarray}
where $U$ represents the internal energy, including thermal and recombination erergy \citep{Hanz1994,Wangc2016a,Klencki2021,Marchant2021}. $\alpha_{\rm th}$ denotes the fraction of the internal energy contributing to envelope ejection. We adopt $\alpha_{\rm th}=0.5$ (Case 1) and $1$ (Case 2). %The chosed values of $\alpha_{\rm th}$ in Case 1 and Case 2 are commonly adapt values in many BPS works. 

For our third treatment (Case 3), we incorporate the $'enthalphy'$ term in the binding energy calculation following \citet{Ivanova2011b}: 
\begin{eqnarray}
  E_{\rm bind} = -\int^{\rm surface}_{\rm core}\left(-\frac{Gm}{r}+U+\frac{P}{\rho}\right){\rm d}m,
  \label{eq:4}
\end{eqnarray}
where $P$ and $\rho$ represent the pressure and density of each mass shell. The quantity $U+P/\rho$ is generally known as $enthalpy$ $(H)$. It should be noted that enthalpy consideration does not introduce new energy sources, but rather redistributes the deposited heat during the spiral-in phase \citep{Ivanova2011b,Ivanova2013}. Following \citet{Ivanova2011a}, we define the core-envelope boundary as {the point with the maximum compression $P/\rho$ inside the hydrogen burning shell} (characterized by the peak values of the local sonic velocity; see also \citealt{Marchant2021,Vigna2022}), {a choice motivated by the need to prevent the donor remnant from expanding immediately after CE ejection.} 

\begin{figure}
    \centering
    \includegraphics[width=\columnwidth]{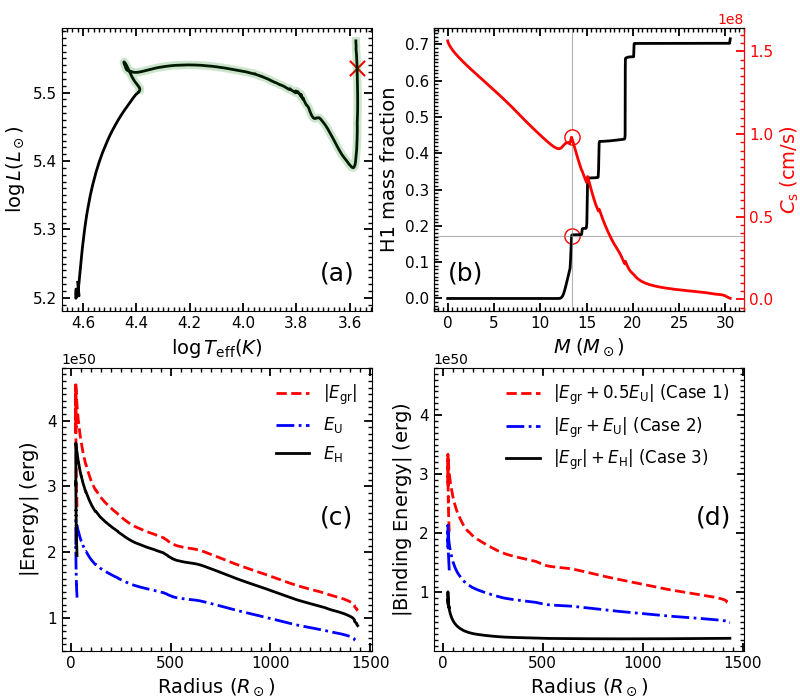}
    \caption{Binding energy calculation example for a $34\,M_\odot$ primary star. Panel (a): The evolutionary track in the HR diagram with the red cross indicating the specific evolutionary stage analyzed. {The thick green line indicates that the star contains a He core.} Panel (b): Radial profiles of hydrogen mass fraction ($X_H$) and the {sonic velocity} $C_{\rm s}\equiv \sqrt{P/\rho}$ at this stage, where open circles mark the core-envelope boundary. Panel (c): The evolution of the primary envelope energy components (gravitational energy $E_{\rm gr}$, internal energy $E_{\rm U}$, and enthalpy term $E_{\rm H}$), {as functions of stellar radius at each specific evolutionary stage (covering the post-MS stages)}. Panel (d): {Similar to panel (c)}, but for the integrated envelope binding energy vs. radius for the three cases.}
    %The sdO first expands towards high luminosity due to the He-shell and H-shell burning.} 
    \label{fig:2}
\end{figure}

Figure~\ref{fig:2} illustrates our binding energy calculation methodology for an initial $34\,M_\odot$ star. Panel (a) shows the evolutionary track in the HR diagram, with the red cross marking the specific evolutionary stage analyzed in panel (b). In panel (b), we present the radial profiles of hydrogen (H) mass fraction and {the sonic velocity} $C_{\rm s}\equiv \sqrt{P/\rho}$, where we identify the core-envelope boundary at the $C_{\rm s}$ extremum ($X_{\rm H}\approx 0.18$), yielding a core mass of $\sim 13.5\,M_\odot$. {Panel (c) shows the evolution of the primary envelope energy components, gravitational energy ($E_{\rm gr}$), internal energy ($E_{\rm U}$) and enthalpy ($E_{\rm H}$), as functions of stellar radius, where the radius corresponds to each specific evolutionary stage, covering the stages following the formation of a He core, i.e., post-MS.} The resulting binding energies for all three cases appear in panel (d). We see that Cases 1 and 2 exhibit a decreasing binding energy with stellar expansion, while Case 3 maintains a relatively constant binding energy through out the late expansion phase, which is similar to what was observed by \citet{Wangc2016a}.  

\textbf{Step 2 - Determining Post-CE Binary Parameters ($a_{\rm post-CE}$, $M_2$):} We conduct a series of binary star evolutionary calculations for BH + MS systems (Stage 6), screening the parameter space to identify models that match observational BHXB data within uncertainties (similar methods as done by \citealt{Willems2005,Fragos2009,Wong2012,Wong2014,Sorensen2017}). For each viable model in the parameter space, we determine the BH mass ($M_{\rm BH}$), companion mass ($M_2$), and orbital separation ($a_{\rm s6}$) at Stage 6, then derive the corresponding post-CE (Stage 4) parameters. A system is considered successfully reproduced when its evolutionary track simultaneously matches the three observed parameters ($M_{\rm BH}$, $M_2$, and orbital period $P_{\rm orb}$) within their error ranges. To account for theoretical uncertainties in period predictions arising from imperfect knowledge of angular momentum loss mechanisms, mass loss processes, and tidal interactions \citep[see e.g.,][]{Tauris2006,Postnov2014}, we adopt a conservative $50\%$ uncertainty in the modeled orbital periods relative to the observed values.

From Stage 6 to Stage 5, the binary parameters are affected by the SN explosion event during the BH formation. Our default model assumes zero BH natal kick for simplicity. Under this framework, we apply the Blaauw-Boersma formalism \citep{Blaauw1961,Boersma1961} to describe the instantaneous mass loss during the SN event while neglecting any natal kick to the BH remnant. {Assuming that the sudden mass loss does not affect the binary position and relative velocity at the moment of SN event, the eccentricity $e$ induced by the mass loss and the new semi-axis $a_{\rm p}$ just after the SN event are given by $e=(M_{\rm pre-SN}-M_{\rm BH})/(M_{\rm BH}+M_2)$; $a_{\rm p}=a_{\rm s5}/(1-e)$ \citep{Dewey1987,Verbunt1990}, where $a_{\rm s5}$ is the binary separation at Stage 5. We futher assume that the orbit is rapidly circularized after the SN event \citep{Verbunt1990}, based on the angular momentum conservation, the circularized orbit satisfies $a_{\rm s6}=(1-e^2)a_{\rm p}$. Consequently, we obtain} 
\begin{eqnarray}
  \frac{a_{\rm s6}} {a_{\rm s5}} = \frac{M_{\rm pre-SN}+M_2}{M_{\rm BH}+M_2}.
  \label{eq:5}
\end{eqnarray}
The pre-SN He star mass $M_{\rm pre-SN}$ relates to the initial He star mass $M_{\rm He}$ (equivalent to the post-CE core mass $M_{\rm 1c,s3}$) through the empirical relation derived from detailed stellar evolution models \citet{Woosley2019}:
\begin{eqnarray}
  M_{\rm pre-SN} = 0.463\,M_{\rm He} + 1.49\,M_\odot.
  \label{eq:6}
\end{eqnarray}

From Stage 5 to Stage 4, the orbit evolution is governed by mass loss through the stripped He star's stellar wind, which we consider as a fast (several $1000\;\rm km/s$; \citealt{Vink2022}) and isotropic spherically-symmetric wind that does not interact with the companion star. The assumptions are well-justified for Wolf-Rayet stars where the wind velocity significantly exceeds the typical orbital velocity ($\sim 100\;\rm km/s$), making wind accretion negligible \citep{ElMellah2019,lizw2026}. The change in binary separation due to the wind mass loss then is given by \citep{Postnov2014}
\begin{eqnarray}
  \frac{a_{\rm s5}}{a_{\rm post-CE}} = \frac{M_{\rm He}+M_2}{M_{\rm pre-SN}+M_2}.
  \label{eq:7}
\end{eqnarray}

For each set of binary parameters at Stage 6 that successfully reproduces the observed BHXB properties, we can systematically reconstruct the corresponding post-CE binary configuration at Stage 4. {Using Equation (\ref{eq:1}) together with the binding energy $E_{\rm bind}$ calculated for a given BH progenitor across different evolutionary stages,} we determine the required CE efficiency parameter $\alpha_{\rm CE}$ for each viable solution.

It should be noted that these calculations adopt the Blaauw-Boersma formalism, which neglects natal kicks. The potential impact of non-zero kick velocities on our results will be examined in Section \ref{sec:3.2}.

%\subsection{Nummerical simulation inputs}
\subsection{The BH progenitors}
\label{sec:2.4}
%\subsection{Black hole progenitors}
The formation of BH is strongly linked to pre-SN stellar structures and explosion mechanisms, with current simulations revealing a complex, non-linear relationship between progenitor mass and final BH mass \citep{Ugliano2012, Janka2012, Janka2016, Ertl2016, Sukhbold2016}. This complexity is further compounded in binary systems, where interactions significantly alter pre-SN structures and consequently affect SN outcomes \citep{Schneider2021, David2021, Laplace2021, Schneider2023, Aguilera2023, Schneider2024, Laplace2025, Schneider2025}. Recently, \citet{Schneider2021} combined detailed pre-SN simulations with the parametric explosion model of \citet{Muller2016} to predict stellar fates while accounting for binary interactions with three distinct cases of envelope stripping: Case A (during core H burning), Case B (during H-shell burning), and Case C (after core He burning). Rather than simulating SN explosions directly, we utilize the results of \citet{Schneider2021} to determine whether a star collapses to form a BH (see also Section~\ref{sec:2.5}). 

We assume the CE phase occurs when the donor star has developed a He core but before core He ignition (Case B stripped stars; {following He ignition in the core, the stellar radius undergoes rapid shrinkage.} \citealt{Klencki2021}). This assumption of CE occurrence before maximum radial expansion is observationally justified when tidal effects are not too strong (e.g., \citealt{Niej2017}; see Section \ref{sec:4} for further discussion). According to \citet{Schneider2021}, Case B stripped stars produce BHs through three distinct mass ranges: $\sim 32-40\,M_\odot$, $\sim 46-60\,M_\odot$, and $\sim 70-100\,M_\odot$. The most massive progenitors ($70-100\,M_\odot$) undergo direct collapse, typically forming BHs exceeding $10\,M_\odot$, which are heavier than those in our sample (Table \ref{tab:1}). Lower-mass progenitors $33-60\,M_\odot$) generally form lighter BHs via fallback accretion. In the $32\,M_\odot$ case, although direct collapse occurs due to enhanced CO core compact, the resulting BH mass remains below $\sim 8\,M_\odot$, depending on gravitational binding energy release. Based on these findings, we constrain our analysis to progenitor masses of 32-40$\,M_\odot$ and 46-60$\,M_\odot$ for the three BHXBs.

\subsection{The input parameters}
\label{sec:2.5}

We perform all stellar evolution calculations using the state-of-the-art stellar evolutionary code {Modules for Experiments in Stellar Astrophysics} ({\tt{MESA}}; version 12115; \citealt{Paxton2011,Paxton2013,Paxton2015,Paxton2018,Paxton2019,Jermyn2023}), adopting solar metallicity (Z=0.0142) with initial He fraction Y=0.2703 \citep{Asplund2009} for non-rotating massive star models ({as the strong stellar winds in massive stars cause a marked spin-down; e.g., \citealt{Nathaniel2025}}). Our simulations incorporate Type II opacities \citep{Iglesias1996} and the approx21$\_$cr60$\_$plus$\_$co56.net nuclear reaction network \citep{Ferguson2005}, with convection parameters including a mixing length of $\alpha_{\rm mlt}=1.8$, core overshooting of 0.2 pressure scale heights during H/He burning phases \citep{Stancliffe2015}, and semi-convection efficiency $\alpha_{\rm sc}=0.1$ \citep{Choi2016}. Stellar wind mass loss follows the prescriptions of \citet{Marchant2016}. These chosen input parameters are closely similar to those employed by \citet{Schneider2021}, ensuring consistency when applying their results to determine BH progenitor masses. %while properly accounting for the combined effects of stellar winds, nuclear burning, and mixing processes on pre-SN stellar structures.

%The formation mechanism of a BH largely depends on the pre-SN structure and st
Our binary evolution simulations focus specifically on the formation of BHXBs from BH + MS systems. In this framework, we model the BH as a non-rotating point mass while evolving the MS star using the same physical prescriptions described previously. The BH accretion is limited by the Eddington accretion rate, i.e.,
\begin{eqnarray}
  \dot{M}_{\rm Edd} &=& \frac{4\pi G M_{\rm BH}} {\kappa c \eta} \\
  &\simeq& 2.6\times 10^{-7}M_\odot {\rm yr^{-1}} \left(\frac{M_{\rm BH}}{10M_\odot}\right)\left(\frac{\eta}{0.1}\right)^{-1}\left(\frac{1+X}{1.7}\right)^{-1}, \nonumber
\end{eqnarray}
where $\kappa (\equiv 0.2(1+X)\rm cm^2 g^{-1})$ is the electron scattering opacity \citep{Kippenhahn1990}, $c$ is the speed of light, $\eta$ is the accretion efficiency, which is approximately given by \citep{Podsiadlowski2003} 
\begin{eqnarray}
  \eta = 1-\sqrt{1-\left(\frac{M_{\rm BH}}{3M_{\rm BH,0}}\right)},
\end{eqnarray}
where $M_{\rm BH,0}$ and $M_{\rm BH}$ are the initial and the present mass of the BH, respectively. %The spin evolution of the BH due to the mass accumulation follows same assumption in \citet{Podsiadlowski2003}. % (see also Bardeen1970, Thorne1974, King 1999). 
The unprocessed material is assumed to carry away the specific angular momentum of the BH as irradiation wind \citep{Postnov2014}.  

In our binary evolution calculations, we systematically explore the parameter space by varying three key parameters: (1) the initial BH mass at Stage 6, which we sample from $3.0\,M_\odot$ upwards in increments of $0.3\,M_\odot$ until exceeding the observed values; (2) the companion star mass $M_2$, ranging from the lower observational limit to the maximum value capable of reproducing the observed systems, also in $0.3\,M_\odot$ steps; and (3) the initial orbital period at Stage 6, covering $0.6-5.0$ days with an interval of 0.2 days for GRO J1655 and SAX J1819, and a finer $0.1$ day stepping for 4U 1543 to account for its tighter observational constraints. %This gridding approach ensures comprehensive coverage of the parameter space while maintaining computational efficiency, with the variable resolution adapted to each system's specific requirements.

\section{Results}
\label{sec:3}
\subsection{Binary Grids for BHXBs}
\label{sec:3.1}

\begin{figure}
    \centering
    \includegraphics[width=\columnwidth]{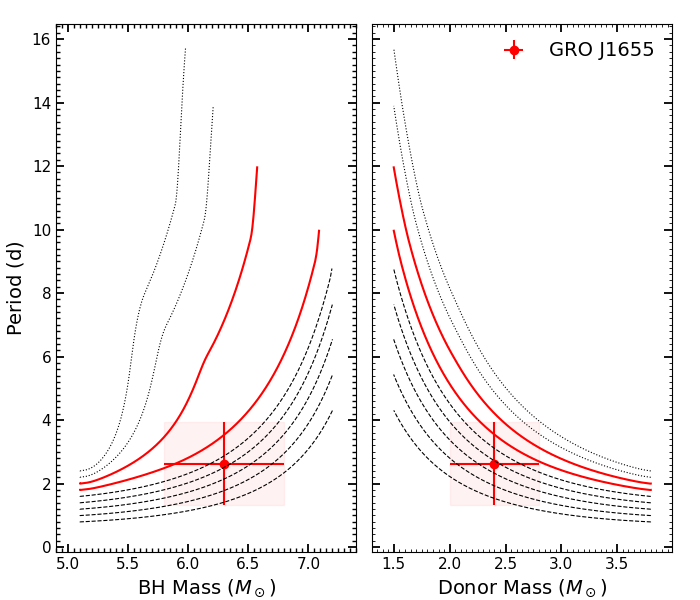}
    \caption{Binary evolutionary tracks for GRO J1655. The initial binary systems have $M_{\rm BH}=5.1\,M_\odot$, $M_{\rm 2}=3.8\,M_\odot$, orbital periods spanning from $0.8$ to $2.4\;\rm d$ in a step of $0.2\;\rm d$. Two classes of solutions successfully reproduce the observed $M_{\rm BH}$, $M_{\rm donor}$, and $P_{\rm orb}$: (1) forbidden configurations (black dashed lines) where the MS star would fill its Roche lobe during post-CE evolution (Stage 4 in Figure~\ref{fig:1}), and (2) physically allowable solutions (red solid lines) that avoid this constraint. The black dotted lines represent binary configurations that fail to reproduce GRO J1655's observed parameters.} 
    \label{fig:3}
\end{figure}

Figure 3 presents a representative set of evolutionary tracks for GRO J1655, where we fix the initial BH mass at $5.1\,M_\odot$ and the secondary mass at $3.8\,M_\odot$ while varying the orbital period from $0.8\;\rm d$ to $2.4\rm d$ in a step of $0.2\;\rm d$. During the mass transfer phase, both the orbital periods and BH masses increase as the donor star loses mass. Among these models, the first seven binaries (black dashed and red solid lines) successfully traverse the observational parameter space of GRO J1655, while systems with longer initial periods (dotted lines) fail to reproduce the observed system due to Eddington-limited accretion \citep{Podsiadlowski2003}. Further examination shows that five of the initially viable cases (black dashed lines) are excluded because their post-CE orbital separations (Stage 4 in Figure~\ref{fig:1}) would cause the secondary to fill its Roche lobe prematurely. Consequently, only the two evolutionary tracks shown as red solid lines represent physically consistent solutions that simultaneously satisfy the observational constraints for GRO J1655. 

\begin{figure}
    \centering
    \includegraphics[width=\columnwidth]{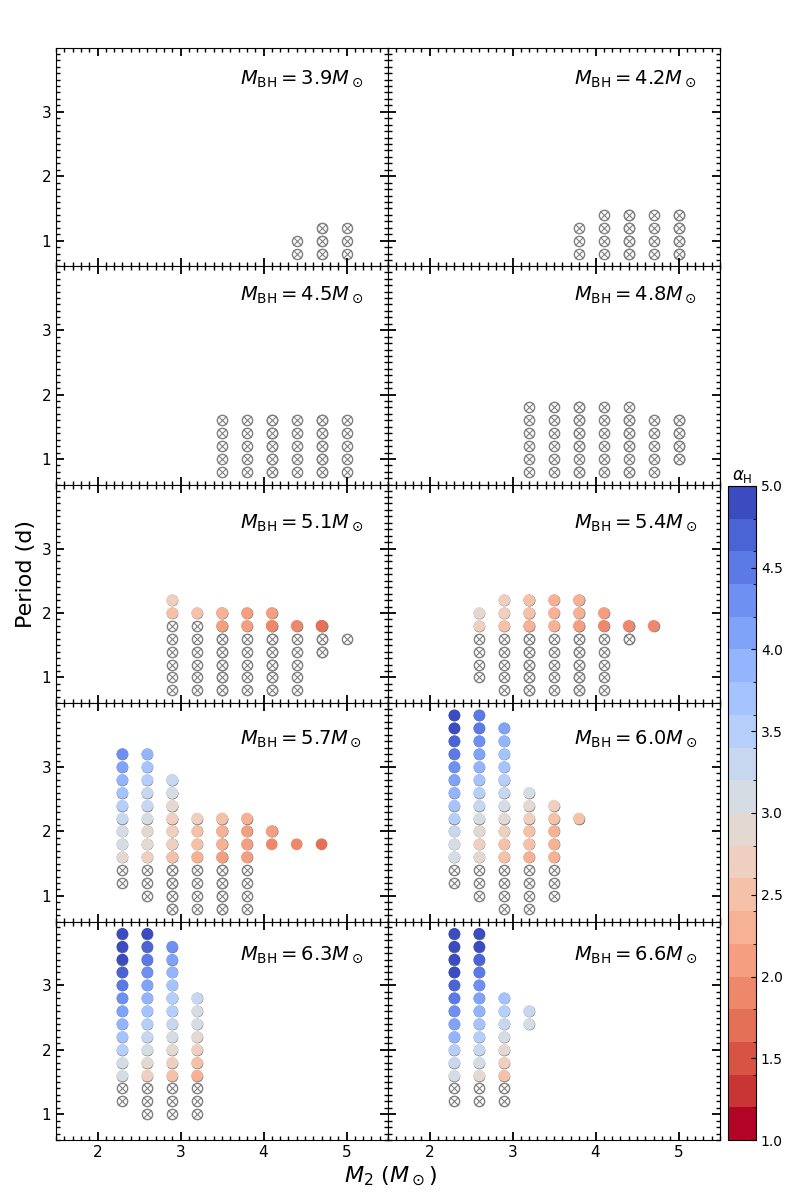}
    \caption{Parameter spaces for GRO J1655. All marked binaries can produce BHXBs consistent with GRO J1655's observed parameters. The viable parameter space consists of: colored solid circles representing allowable solutions where post-CE separations (Stage 4) satisfy $R_2 < R_{\rm 2,L}$, with colors indicating $\alpha_{\rm H}$ values (assuming a $34\,M_\odot$ BH progenitor at $R_1=1000\,R_\odot$); and open circles with crosses denoting forbidden grids violating this Roche lobe condition.}
    %The sdO first expands towards high luminosity due to the He-shell and H-shell burning.} 
    \label{fig:4}
\end{figure}

\begin{figure}
    \centering
    \includegraphics[width=\columnwidth]{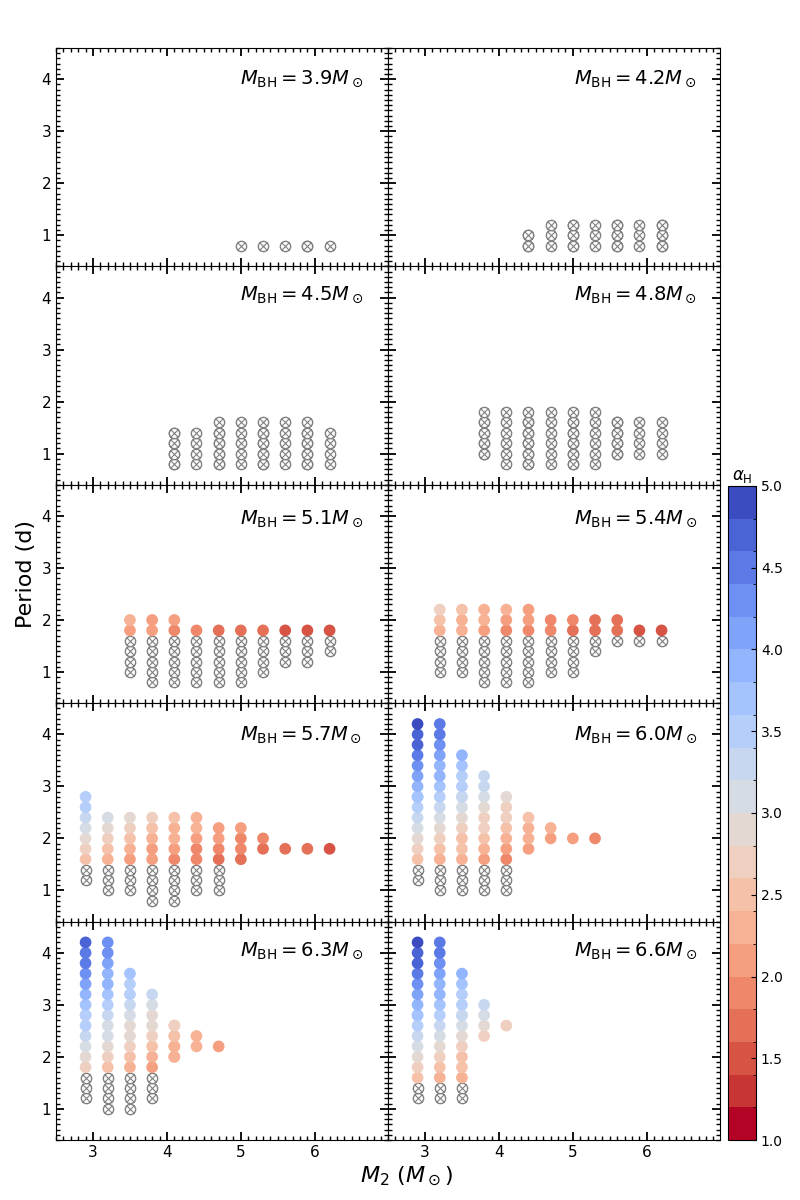}
    \caption{Similar to Figure~\ref{fig:4} but for the parameter spaces of SAX J1819.}
    %The sdO first expands towards high luminosity due to the He-shell and H-shell burning.} 
    \label{fig:5}
\end{figure}

\begin{figure}
    \centering
    \includegraphics[width=\columnwidth]{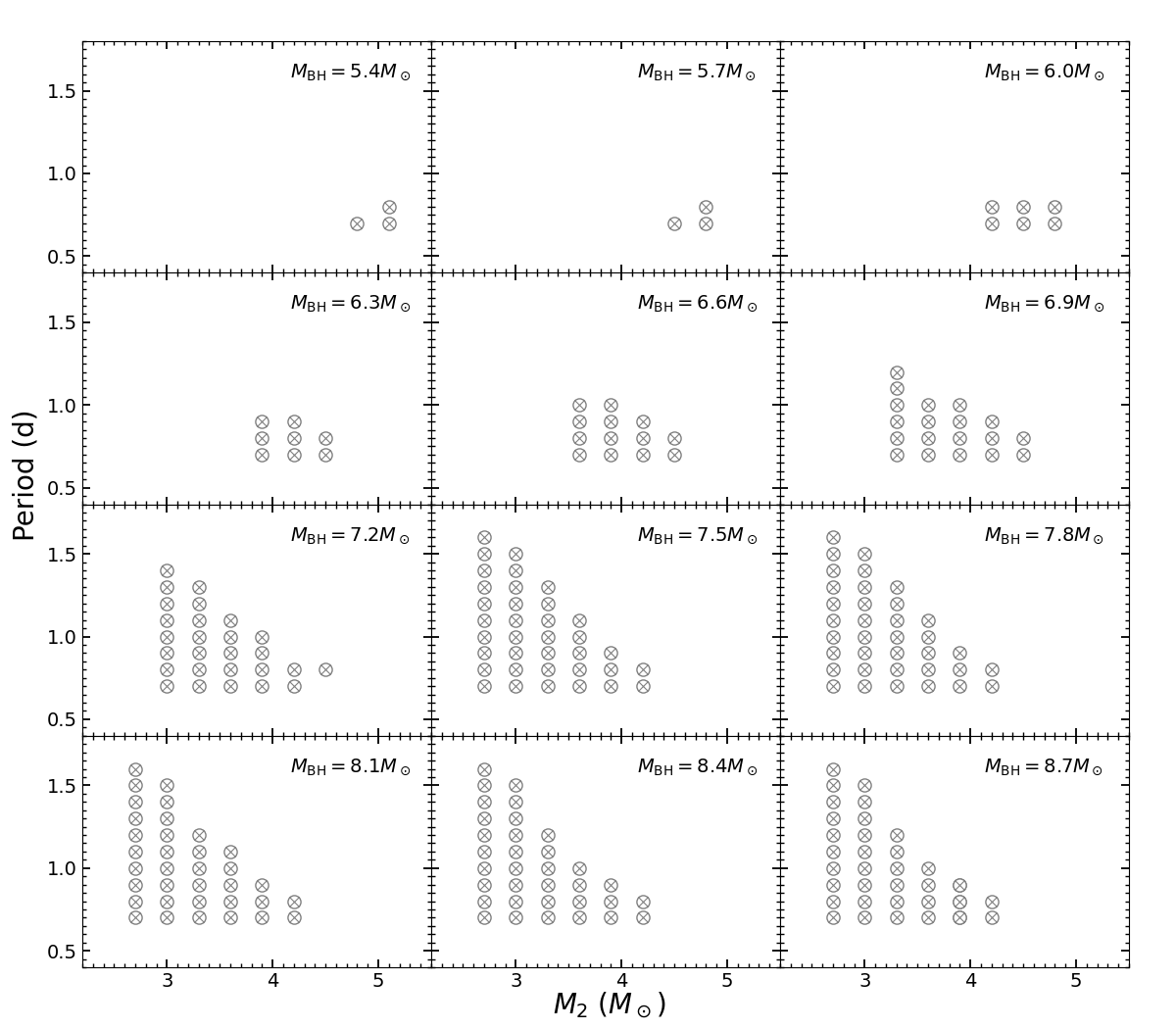}
    \caption{Similar to Figure~\ref{fig:4} but for the parameter spaces of 4U 1543. Notably, no allowable parameter space exists for this system under the default zero-kick model.}
    %The sdO first expands towards high luminosity due to the He-shell and H-shell burning.} 
    \label{fig:6}
\end{figure}

Our complete simulation results\footnote{We adopt a cutoff of $8.7\,M_\odot$ on the BH mass in our grid spaces for 4U 1543, as preliminary tests show that grids with higher BH masses are forbidden grids. When BH natal kicks are included, the allowable grids are further restricted to initial BH masses below $6.3\,M_\odot$ (see Figure~\ref{fig:8} below). Thus, additional simulations with a more massive initial BH would not impact our conclusions. We also apply secondary mass cutoffs of $5.0\,M_\odot$ for GRO J1655 and $6.2\,M_\odot$ for SAX J1819 based on spectral observations, with detailed justification provided in Appendix~\ref{sec:appA}.} for GRO J1655, SAX J1819, and 4U 1543 are presented in Figures~\ref{fig:4}-\ref{fig:6}, respectively. These figures distinguish between two distinct categories of evolutionary outcomes: (1) Forbidden grid points (represented by black open circles with crosses) correspond to cases where the secondary star would prematurely fill its Roche lobe following CE ejection, analogous to the excluded black dashed trajectories in Figure~\ref{fig:3}; and (2) Allowable grid points (shown as colored circles) that successfully reproduce all observed parameters of the respective BHXB systems.

For each allowable grid point, we first obtain the binary parameters of the BH + MS binary at Stage 6 in Figure 1, then derive the corresponding He star + MS binary parameters at Stage 4. Using the known progenitor properties (including envelope binding energy, core mass, and initial separation) for a given evolutionary stage, we calculate the required CE efficiency $\alpha_{\rm CE}$ by applying these parameters to Equation (\ref{eq:1}). The color bars in Figures~\ref{fig:4}-\ref{fig:5} display the derived $\alpha_{\rm H}$ values for a representative $34\,M_\odot$ progenitor with radius $R = 1000\,R_\odot$. Notably, our analysis reveals: (1) no physically allowable solutions exist for 4U 1543 (Figure~\ref{fig:6}); (2) for GRO J1655 and SAX J1819, $\alpha_{\rm H}$ shows a positive correlation with $P_{\rm orb}$, i.e., wider initial orbits (Stage 6) require greater CE efficiencies to unbind the envelope; and (3) at fixed BH mass, $\alpha_{\rm H}$ decreases with increasing companion mass, as predicted by Equation (\ref{eq:1}). These results assume zero BH natal kick; significant kick velocities would modify both the viable parameter space and derived CE efficiencies, as we explored in Section~\ref{sec:3.2}.

%One, kick. Two, other channels. The forbidden and allowable grids are defined only for the Blauuw kick, BH with large BH kick velocity still has a certain probability to produce BH + MS binary for the forbidded grids. 

\subsection{Effect of kick}
\label{sec:3.2}

In our default model, we assume the BH forms with zero natal kick velocity. Under this assumption, the sudden mass loss during the SN explosion invariably widens the orbital separation (from Stage 5 to Stage 6 in Figure~\ref{fig:1}), allowing the post-CE binary separations to be directly derived from the parameter spaces shown in Figures~\ref{fig:4}-\ref{fig:6}. However, the scenario where BHs receive substantial natal kicks cannot be entirely ruled out, as suggested by observational evidence from some BH binaries (e.g., \citealt{Tauris1999,Repetto2015,Mirabel2017,Zhaoy2023,Dashwood2024,Burrows2025,Nagarajan2025,Abdulghani2025}). If a BH forms with a significant kick, {the random orientation of the kick can either increase or decrease the orbital separation}, thereby modifying the allowable parameter space to some degree, and consequently affecting the inferred CE efficiencies. In this section, we examine how BH natal kicks influence our previous results.

The modifications to the orbital parameters resulting from BH natal kicks are computed following the methodology of \citet{Hurley2002}, as outlined below. We assume that the binary orbit is circular prior to BH formation (stage 5 in Figure~\ref{fig:1}) and that the kick is instantaneously imparted to the newborn BH. Based on the Keplerian orbital equations and the principle of angular momentum conservation, the post-kick semimajor axis $a_{\rm n}$ and eccentricity can be derived as follows:
\begin{eqnarray}
  \frac{1}{a_{\rm n}} = \frac{2}{a_{0}} - \frac{GM_{\rm n}}{V^2_{\rm n}},
  \label{eq:8}
\end{eqnarray}
and 
\begin{eqnarray}
  1-e^2_{\rm n} = \frac{|\vec{r}\times \vec{V}_{\rm n}|^2}{GM_{\rm n}a_{\rm n}},
  \label{eq:9}
\end{eqnarray}
where the subscripts $0$ and $n$ denote the binary parameters before and after the kick, respectively ($M_{\rm n} = M_{\rm BH}+M_2$). The vector $\vec{V}_{\rm n}$ represents the relative velocity after the kick ($V_{\rm n} = |\vec{V}_{\rm n}|$), which can be calculated by $\vec{V}_{\rm n} = \vec{V}_{\rm orb} + \vec{V}_{\rm K}$, where $\vec{V}_{\rm orb}$ is the pre-SN orbital velocity. $\vec{r}$ is the initial separation vector between the two stars. In this study, we adopt the simplifying assumption that the kick direction is randomly distributed. Thus, for a given kick speed $V_{\rm K}$, the post-kick orbital parameters can be determined using Equations~(\ref{eq:8}-\ref{eq:9}).  

\subsubsection{Formation of 4U 1543 with a natal kick}
\label{sec:3.2.1}
As discussed above, no viable parameter space exists for the formation of 4U 1543 if the BH forms with zero natal kick velocity. This is because the orbital separations at Stage 4 for binaries within the parameter space of Figure~\ref{fig:6} are sufficiently small that the MS companions would fill their Roche lobes, preventing the system from evolving into an X-ray binary. However, if the BH receives a natal kick, the post-SN orbital separation may decrease depending on the kick direction. To examine this scenario, we adopt a BH progenitor of $34\,M_\odot$ undergoing CE evolution at $1000\,R_\odot$ (Initial parameter choices affect only CE efficiency estimates without impacting our conclusions). We define a minimum separation $a_{\rm min}$ at Stage 4 where the MS star just fills its Roche lobe post-CE, determined by the secondary's mass and radius. For a given kick velocity and each grid point in Figure~\ref{fig:6}, we generate $10^5$ binaries with orbital separations at Stage 4 distributed logarithmically between $a_{\rm min}$ and $10a_{\rm min}$. Finally, we count systems that successfully evolve into 4U 1543-like BHXBs, assuming uniform probability per grid point.

\begin{figure}
    \centering
    \includegraphics[width=\columnwidth]{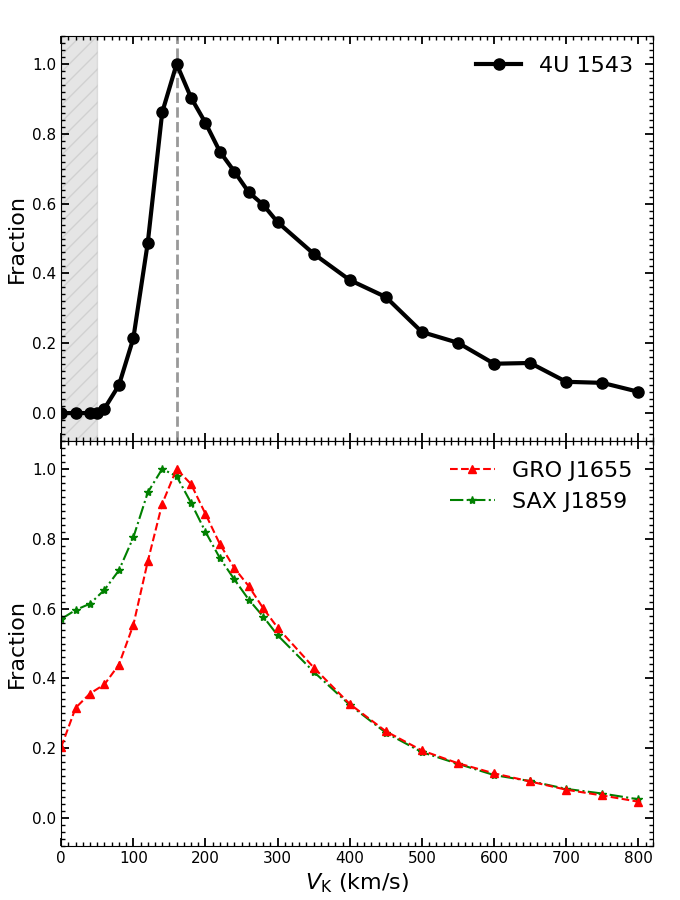}
    \caption{Upper panel: The fraction of the sampled binary systems that successfully evolve into the parameter space corresponding to 4U 1543. Lower panel: Similar to the upper panel, but for GRO J1655 (red dashed line) and SAX J1819 (green dash-dot line). For a given kick velocity, we randomly generate $10^5$ He star + MS binaries with varying orbital separations (Stage 4 in Figure~\ref{fig:1}), and count systems that subsequently fall within the target parameter space. The resulting fractions are normalized to their peak value. The hatched region in the upper panel indicates kick velocities ($V_{\rm K}\leq 50\;\rm km/s$) for which no binary systems can fall within the parameter spaces in Figure~\ref{fig:6}, representing the minimum kick velocity required for the formation of 4U 1543.}
    \label{fig:7}
\end{figure}

The statistical results, normalized to their peak value, are shown the upper panel of Figure~\ref{fig:7}. Our analysis reveals that for kick velocities $V_{\rm K} \leq 50 \rm km/s$, no binary systems fall within the parameter space required for 4U 1543 formation as shown in Figure~\ref{fig:6}. This strongly suggests that if 4U 1543 formed through isolated binary evolution, its BH must have received a natal kick of at least $50 \;\rm km/s$. We identify a preferred kick velocity of $V_{\rm K} = 160 \;\rm  km/s$, which maximizes the probability of forming 4U 1543-like BHXBs. Furthermore, the formation probability becomes significantly reduced for systems with extremely high kick velocities ($V_{\rm K} \gtrsim 800 \;\rm km/s$). For comparison, we also calculated the corresponding results for GRO J1655 and SAX J1819, which are shown in the lower panel. Similarly, the preferred kick velocity that maximizes the formation probability for GRO J1655/SAX J1819-like BHXBs is about $140$–$160 \;\rm km/s$. We note, however, that the above conclusions should be interpreted with caution, owing to the small sample size of observed BHXBs and the incomplete understanding of the physical mechanisms governing natal kicks.
%The statistical results are presented in Figure~\ref{fig:9}, where the number is normalized upon the peak value. For $V_{\rm K}\leq 50\;\rm km/s$, no binary can fall into the parameter spaces in Figure~\ref{fig:5}. It has important implications that if 4U 1543 is formed through isolated binary evolution, the BH needs to be born with a kick velocity no less than $50\;\rm km/s$. We also find the prefered BH kick velocity is $V_{\rm K}=160\;\rm km/s$ of which there is a larger possibility to form 4U 1543-like BHXBs. Futhermore, the possibility is small for BH born with extremely high kick velocity of $V_{\rm K}\gtrsim 800\;\rm km/s$. 

\subsubsection{Influence of kick on BH born mass}
\label{sec:3.2.2}

\begin{figure}
    \centering
    \includegraphics[width=\columnwidth]{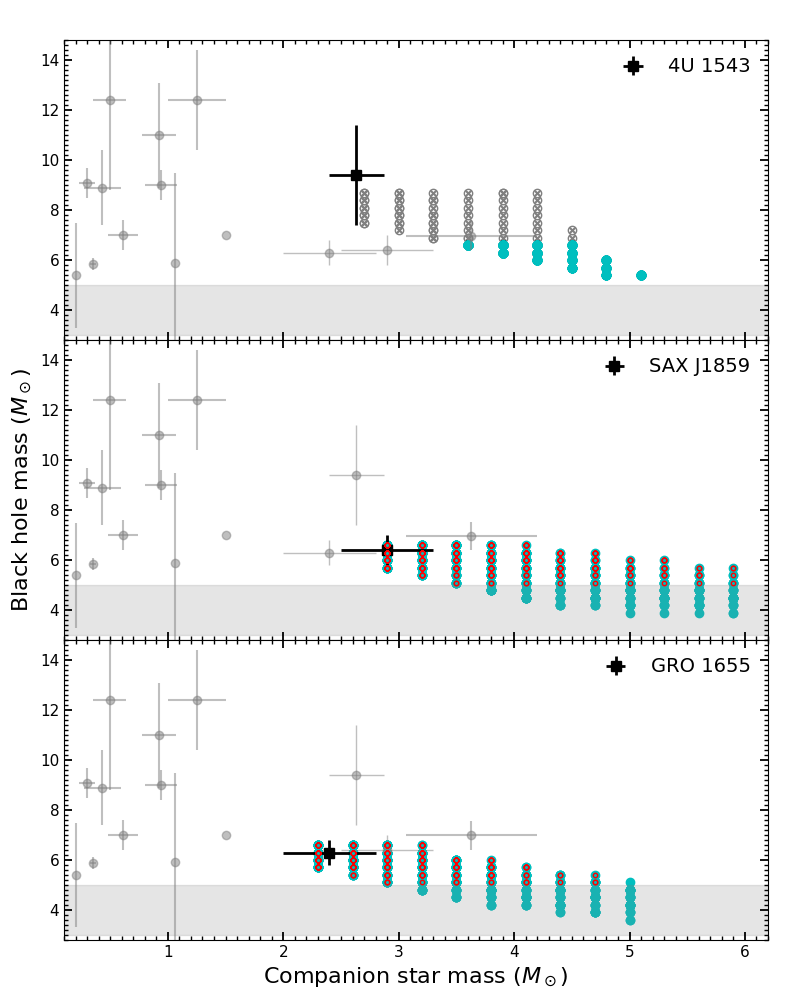}
    \caption{The BH mass and companion star mass distributions for the grid points shown in Figures~\ref{fig:4}-\ref{fig:6}. The upper panel distinguishes between the forbidden configurations (gray open circles with crosses) and viable systems when BH natal kicks are included (cyan circles). The middle and lower panels respectively display the allowable parameter space for systems excluding (red open circles) and including (cyan circles) BH natal kicks. The observed BHXB samples are taken from \citet{Corral2016}.} 
    \label{fig:8}
\end{figure}

The influence of BH natal kicks extends beyond 4U 1543 to affect the parameter spaces of SAX J1819 and GRO J1655 as well. Figure~\ref{fig:8} displays the BH mass and companion star mass distributions corresponding to the grid points in Figures~\ref{fig:4}-\ref{fig:6}. The upper panel distinguishes between forbidden grids (gray open circles with crosses) and viable systems when BH kicks are considered (cyan circles). The middle and lower panels compare systems without BH kicks (red open circles, representing allowable grids from Figures~\ref{fig:4} and~\ref{fig:5}) to those with kicks. Notably, BH natal kicks significantly expand the allowable parameter space. Most remarkably, for both SAX J1819 and GRO J1655, the inclusion of natal kicks allows for BH formation in the so-called mass-gap region $\sim 3-5\,M_\odot$ \citep{Shaoy2022,Wangs2024}, a scenario prohibited in the zero-kick case.

\subsubsection{Influence of kick velocity on CE efficiency}
\label{sec:3.2.3}

As discussed above, when accounting for non-zero BH kick velocities, the binary separation at Stage 4 cannot be determined directly from Equations (\ref{eq:5}-\ref{eq:7}). To address this, we implement a Monte Carlo approach to systematically evaluate how kick velocities influence CE efficiency. We first select representative grid points from Figure~\ref{fig:4} (for GRO J1655) and Figure~\ref{fig:5} (for SAX J1819), each defining the minimum allowable separation $a_{\rm min}$ at Stage 4. For each grid point, we generate $10^6$ binary systems with orbital separations log-uniformly distributed across $(a_{\rm min}, 10a_{\rm min})$, then apply the SN kick and track whether the post-SN parameters fall within the original grid point. The fraction of systems remaining within the grid boundaries serves as a statistical weight for the corresponding CE efficiency calculated via Equation (\ref{eq:1}).

\begin{figure}
    \centering
    \includegraphics[width=\columnwidth]{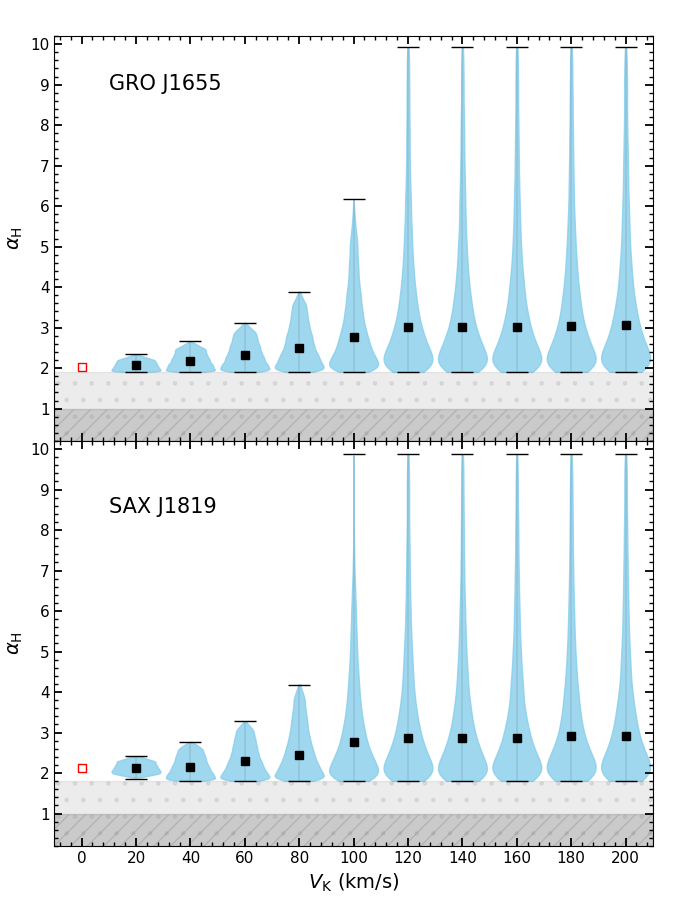}
    \caption{The influence of kick velocity on the CE efficiency ($\alpha_{\rm H}$). The chosen grid points for GRO J1655 and SAX J1819 are $M_{\rm BH}=5.1, M_{\rm 2}=4.1\,M_\odot, P_{\rm orb}=2.0\;\rm d$, and $M_{\rm BH}=5.7, M_{\rm 2,i}=4.4\,M_\odot, P_{\rm orb}=2.0\;\rm d$, respectively. The BH progenitor is assumed to have an initial mass of $32\,M_\odot$ with CE occurring when the progenitor reaches a stellar radius of $1000\,R_\odot$. For the zero-kick case, the corresponding $\alpha_{\rm H}$ values are shown by red open squares. The distribution of $\alpha_{\rm H}$ values is illustrated in the violin plot, where the median values are represented by black squares. A cutoff of $\alpha_{\rm H} = 10$ is applied for systems with large kick velocities. The light gray hatched regions denote cases where the post-CE orbital separation falls below the minimum threshold $a_{\rm min}$, while the dark gray hatched regions indicate scenarios with $\alpha_{\rm H} < 1$.}
    %The sdO first expands towards high luminosity due to the He-shell and H-shell burning.} 
    \label{fig:9}
\end{figure}

Figure~\ref{fig:9} demonstrates the impact of kick velocity on the CE efficiency parameter $\alpha_{\rm H}$ for selected systems: GRO J16550 ($M_{\rm BH}=5.1 \,M_\odot,\; M_2=4.1\,M_\odot,\; P_{\rm orb}=2.0 \rm \; d$) and SAX J1819 ($M_{\rm BH}=5.7 \,M_\odot,\; M_2=4.4\,M_\odot,\; P_{\rm orb}=2.0 \rm \, d$), both originating from a $32\,M_\odot$ progenitor undergoing CE evolution at $R=1000\,R_\odot$. The zero-kick case (red open squares) is also shown for comparison. Our Monte Carlo results are presented by violin plots with median values (black squares). The light grey hatched regions indicate systems where the post-CE separation falls below $a_{\rm min}$, while the dark grey regions correspond to cases with $\alpha_{\rm H}<1$. Our results show that while the estimated CE efficiencies for systems with kicks may be slightly lower than those in the zero-kick case, the median $\alpha_{\rm H}$ values exhibit a gradual increase with higher BH kick velocities. Most significantly, even when accounting for BH kicks, all $\alpha_{\rm H}$ values remain greater than 1 (mainly because of the limitation of $a_{\rm min}$).

\subsection{Progenitor Mass Dependence of Minimum $\alpha_{\rm CE}$}
\label{sec:3.3}

\begin{figure*}
    \centering
    \includegraphics[width=\textwidth]{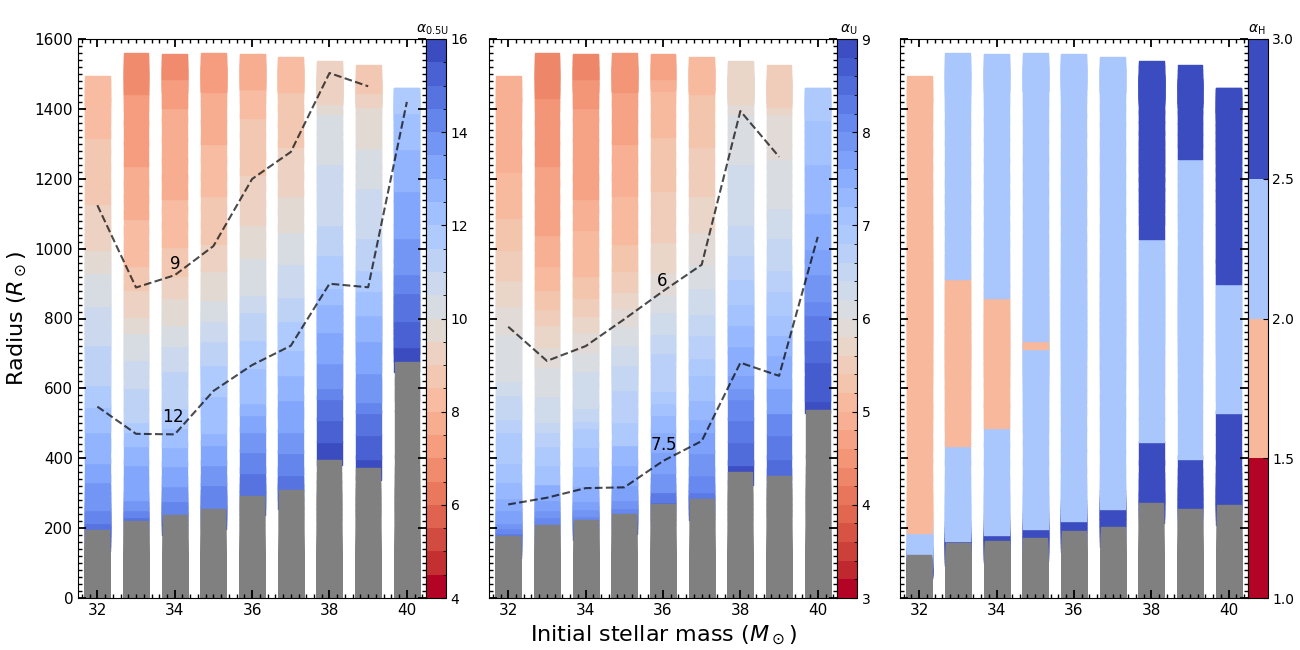}
    \caption{The minimum values of CE efficiency for GRO J1655 vary with the initial primary mass and the stellar radius. The left, middle and right panels correspond to the $\alpha_{\rm 0.5U}$, $\alpha_{\rm U}$ and $\alpha_{\rm H}$ cases respectively. {The shaded gray region indicates CE efficiencies above the upper limit of the colorbar.} Dashed lines indicate constant minimum CE efficiency values. Notably, when enthalpy is included (right panel), the minimum CE efficiencies show significantly less variation, typically confined to a narrow range of $1.5-3$.}
    %The sdO first expands towards high luminosity due to the He-shell and H-shell burning.} 
    \label{fig:10}
\end{figure*}

\begin{figure*}
    \centering
    \includegraphics[width=\textwidth]{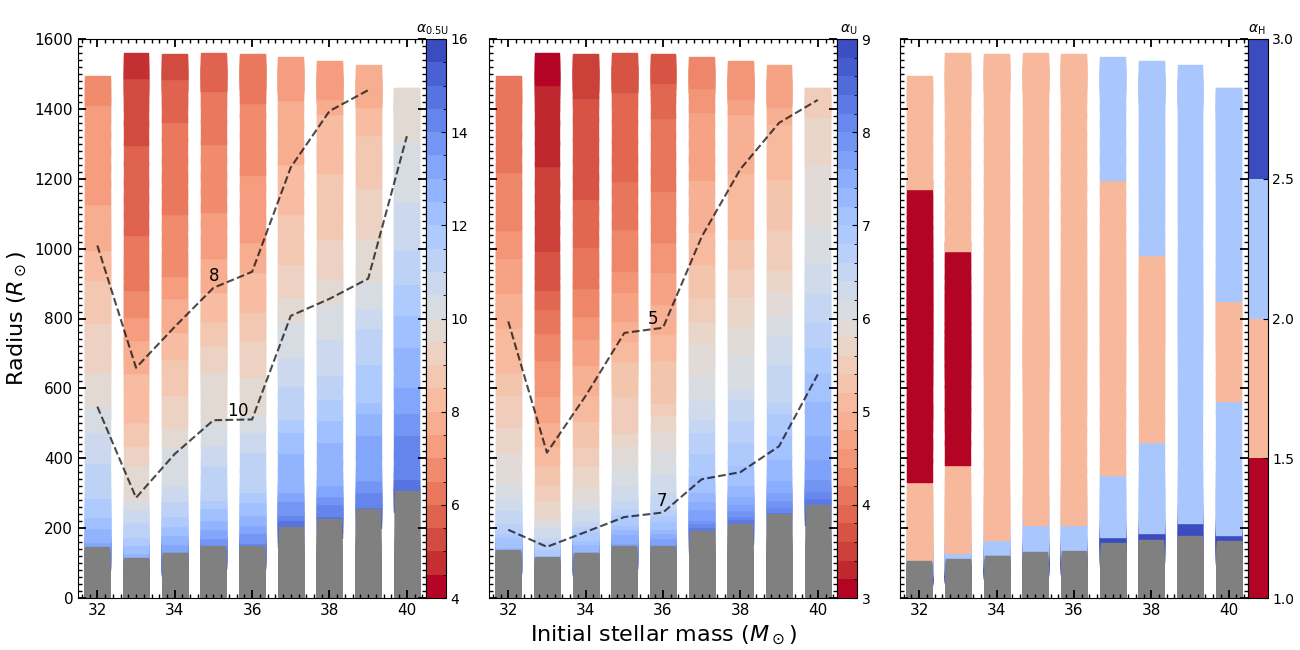}
    \caption{Similar to Figure~\ref{fig:10}, but for the case of SAX J1819.}
    %The sdO first expands towards high luminosity due to the He-shell and H-shell burning.} 
    \label{fig:11}
\end{figure*}

\begin{figure*}
    \centering
    \includegraphics[width=\textwidth]{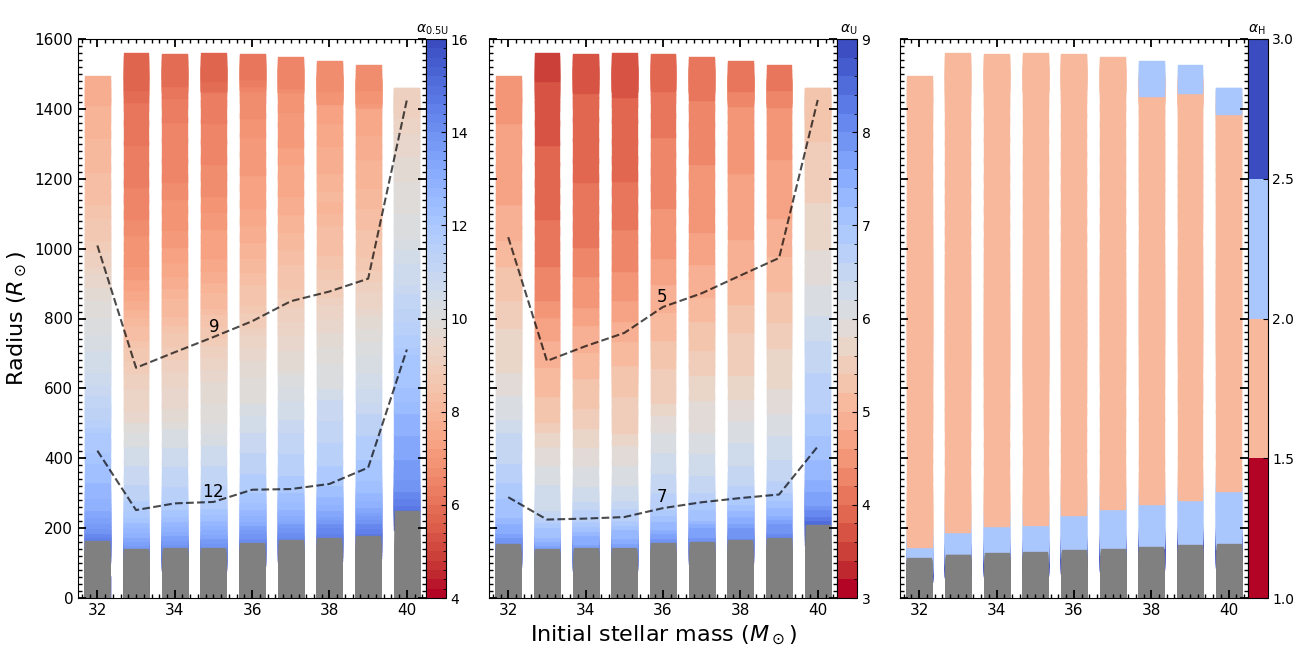}
    \caption{Similar to Figure~\ref{fig:10}, but for the case of 4U 1543, where the minimum CE efficiency values are calculated using the critical orbital separation $a_{\rm min}$ at Stage 4.}
    \label{fig:12}
\end{figure*}

The CE efficiency parameter $\alpha_{\rm CE}$ is primarily determined by three key factors: (1) the initial primary mass (BH progenitor), which sets the stellar structure; (2) the evolutionary stage at CE onset, which affects the envelope binding energy $E_{\rm bind}$; and (3) the post-CE binary parameters that govern the orbital energy release. For each allowable grid point (4U 1543 can also have allowable grid points when accounting for significant BH kick velocities; see Figure~\ref{fig:8}) in our parameter space, we compute the CE efficiency for fixed BH progenitor masses at specific evolutionary stages (characterized by stellar radii), enabling us to determine the minimum required $\alpha_{\rm CE}$ values as functions of initial primary mass and evolutionary stage, as shown in Figure~\ref{fig:10} (GRO J1655), Figure~\ref{fig:11} (SAX J1819) and Figure~\ref{fig:12} (4U 1543). It should be noted that the minimum CE efficiency is primarily governed by the threshold binary separation ($a_{\rm min}$), a quantity unaffected by BH natal kicks. Here we present the results only for BH progenitors with masses of $32-40\,M_\odot$; the corresponding results for more massive BH progenitors can be found in the Appendix~\ref{sec:appB}.

We see that the minimum CE efficiency does not always increase monotonically with the mass of the BH progenitor when other parameters are held fixed. This behavior can be understood as follows. For a BH progenitor filling its Roche lobe at a given evolutionary stage, a more massive progenitor typically produces a more massive He star after CE ejection. According to Equation (\ref{eq:5}-\ref{eq:7}), a more massive He star tends to result in a wider post-BH-formation orbit. Inversely, for a specific grid point in the parameter space, a more massive progenitor would lead to a shorter post-CE orbital separation. Consequently, although a more massive BH progenitor generally has a larger envelope binding energy, the shorter post-CE separation can yield a lower estimated CE efficiency (according to Equation \ref{eq:1}). Furthermore, the allowable parameter space varies with the initial black hole progenitor mass, which affects the derived minimum CE efficiency values. This dependence explains why the 33 $M_\odot$ progenitor exhibits relatively lower minimum CE efficiency compared to the 32 $M_\odot$ case when both systems are undergoing CE evolution at similar evolutionary stages.

When enthalpy is not considered (left and middle panels of Figures~\ref{fig:10}-\ref{fig:12}), the envelope becomes progressively easier to eject at later evolutionary stages due to the decreasing binding energy as the star expands (Figure~\ref{fig:2}). In contrast, when enthalpy is included (right panels), the derived $\alpha_{\rm H}$ values exhibit only modest variations for a given BH progenitor, reflecting the weak dependence of binding energy on stellar radius (Figure~\ref{fig:2}). The derived CE efficiency constraints are: 
\begin{itemize}
\item{GRO J1655: $\alpha_{\rm 0.5U}\gtrsim 7-9 $, $\alpha_{\rm U}\gtrsim 4-6$, $\alpha_{\rm H} \gtrsim 1.7$.}
\item{SAX J1819: $\alpha_{\rm 0.5U}\gtrsim 5-8 $, $\alpha_{\rm U}\gtrsim 3-5$, $\alpha_{\rm H} \gtrsim 1.4$.}
\item{4U 1543: $\alpha_{\rm 0.5U}\gtrsim 5.5-9$, $\alpha_{\rm U}\gtrsim 3.5-5$, $\alpha_{\rm H} \gtrsim 1.6$.}
\end{itemize}

Notably, all three efficiency values exceed unity across these systems.

\subsection{Suggested $\alpha_{\rm CE}$ lower limit}
\label{sec:3.4}
\begin{figure}
    \centering
    \includegraphics[width=\columnwidth]{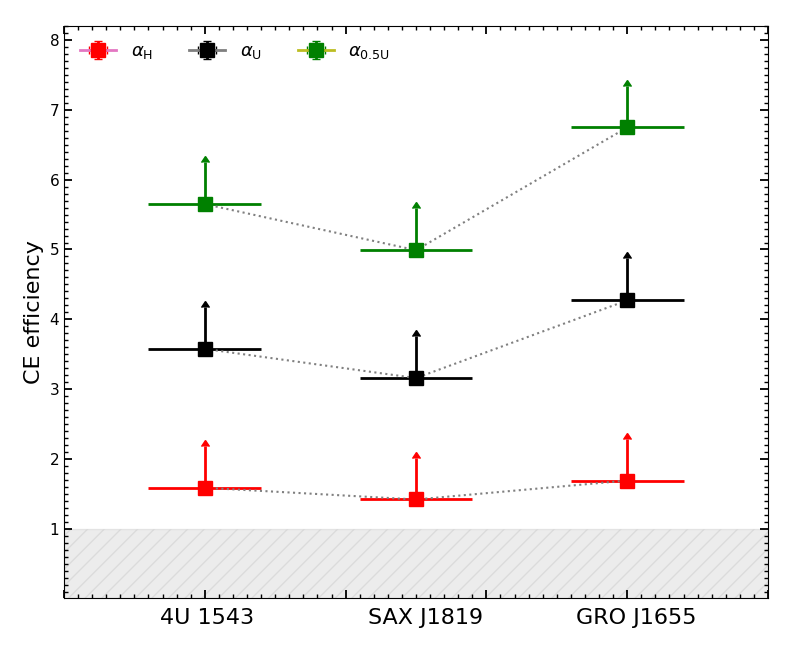}
    \caption{The minimum CE efficiencies required to form the three BHXBs. The red, black, and green symbols correspond to the values of $\alpha_{\rm H}$, $\alpha_{\rm U}$, and $\alpha_{\rm 0.5U}$, respectively. To simultaneously explain the formation of all three BHXBs, higher efficiency thresholds are required: $\alpha_{\rm 0.5U} \gtrsim 6.7, \alpha_{\rm U} \gtrsim 4.2$, and $\alpha_{\rm H} \gtrsim 1.7$.} 
    %The sdO first expands towards high luminosity due to the He-shell and H-shell burning.} 
    \label{fig:13}
\end{figure}

Our results demonstrate that multiple factors significantly influence the estimation of CE efficiencies, including binary progenitor masses, the evolutionary stage of the BH progenitor at CE initiation, and BH natal kicks. Given the substantial uncertainties in these parameters, precise constraints on CE efficiency ranges remain challenging. However, we can establish well-defined lower limits for these efficiencies. Figure~\ref{fig:13} presents the minimum CE efficiencies $\alpha_{\rm H}$, $\alpha_{\rm U}$ and $\alpha_{\rm 0.5U}$ for all three studied BHXBs, determined by their respective $a_{\rm min}$ values, which are unaffected by BH kicks. The dependence of minimum CE efficiency on initial BH progenitor mass is shown in Figure~\ref{fig:C1} (Appendix~\ref{sec:appC}). Crucially, none of the systems permit solutions with CE efficiencies below unity. To simultaneously explain the formation of all three BHXBs, we require significantly higher efficiency thresholds: $\alpha_{\rm 0.5U} \gtrsim 6.7, \alpha_{\rm U} \gtrsim 4.2$, and $\alpha_{\rm H} \gtrsim 1.7$. These findings strongly suggest that successful CE ejection in massive binaries necessitates high efficiency parameters, in agreement with previous studies of massive binary evolution \citep{Podsiadlowski2003,Fragos2019}.

\section{Discussion}
\label{sec:4}
In this study, we have constrained the CE phase in massive binary systems through analysis of three observed BHXBs. Using the standard energy formalism for CE ejection, we have determined robust lower limits for the CE efficiency parameters. {Specifically}, all three systems require CE efficiencies greater than unity $\alpha_{\rm CE} > 1$, even when accounting for enthalpy contributions that substantially reduce the envelope binding energy. {This strongly suggests that either additional energy sources are required within the energy formalism of the commonly adopted CE description, or alternatively, the description itself needs to be modified. We now proceed to discuss our results in more detail.}
%This presents a physical inconsistency, as efficiencies exceeding unity imply insufficient orbital energy to eject the envelope even under maximal energy conversion. Several potential explanations may account for this unexpected result:

\begin{itemize}
  \item \textbf{Core-envelope boundary:} {One of the largest uncertainties in calculating the envelope binding energy lies in the definition of the core-envelope boundary. In this work, we adopt a physically motivated prescription, where the boundary is determined by the maximum compression point \citep{Ivanova2011a}. Under this assumption, the core-envelope boundary typically corresponds to a hydrogen mass fraction of $X_{\rm H} \approx 0.1$–$0.3$. This adopted boundary is located farther out (i.e., closer to the stellar surface) compared to the commonly used criterion in binary population synthesis studies, which often take $X_{\rm H} = 0.1$ as the core-envelope boundary (e.g., \citealt{Kruckow2016,Kruckow2018}). Moving the boundary outward would undoubtedly reduce the envelope binding energy and thus lower the inferred $\alpha_{\rm CE}$ values. However, a more outward boundary also implies that more hydrogen remains in the envelope, which may lead to significant re-expansion of the stripped core after the CE phase (e.g., \citealt{DeMarco2011,Marchant2021,Vigna2022}). Recent simulations by \citet{Vigna2022} examined the stellar response following the removal of the hydrogen-rich envelope. They found that the boundary allowing the star to avoid rapid re-expansion lies above the maximum compression point. Nevertheless, their simulations focused on stars with masses below $25\,M_\odot$, while our work primarily targets more massive stars ($\geq 32\,M_\odot$). This mass difference may result in different post-stripping behaviour. A key distinction is that for stars below $25\,M_\odot$, the maximum compression point is generally located at $X_{\rm H} < 0.1$ \citep{Vigna2022}. In contrast, for stars with masses $\gtrsim 32\,M_\odot$, we find that the maximum compression point falls in the range $0.1 < X_{\rm H} < 0.3$. Therefore, both the appropriate definition of the core-envelope boundary and the post-CE evolution of the stripped core remain open questions that warrant further investigation.}

  \item \textbf{Extra energy sources - Nuclear energy:} {\citet{Podsiadlowski2010} suggested that nuclear energy could contribute to envelope ejection. In this scenario, hydrogen-rich material from the low-mass companion is injected into the helium-burning shell of the massive donor during the late CE phase, triggering a nuclear runaway that explosively ejects both the hydrogen and helium layers. The result is a close binary containing a compact object and a low-mass companion. This channel is, however, very rare and requires that the low-mass companion overflows its Roche lobe during the self-regulated spiral-in phase to transfer hydrogen-rich material deep into the donor's helium shell \citep[see also][]{Ivanova2002,Ivanova2020}. This channel involves complex mass transfer processes within the CE that are difficult to model with current simulations, and therefore remains a topic for future work.}

  \item \textbf{Extra energy sources - Jets:} {The possibility that jets launched from the accreting companion serve as an additional energy source during CE evolution has gained substantial support from recent theoretical, numerical, and observational studies. Numerical simulations have demonstrated that jets can remove mass before and during the CE phase, reduce drag forces, and facilitate envelope unbinding (e.g., \citealt{Soker2004,Soker2015,Shiber2024,Weiner2025}). Furthermore, the newly proposed jetted mass-removal accretion mechanism indicates that MS stars can accrete at high rates, provided that jets remove the outer, high-entropy layers of the accreted envelope (\citealt{Bear2025,Scolnic2025,Cohen2026}; but see \citealt{Zou2022}). Observationally, jets appear to be the most robust observable ingredient of CE evolution, found in about $40\%$ more post-CE planetary nebulae than dense equatorial outflows \citep{Soker2025}. Collectively, these developments suggest that jets may be regarded as a primary candidate for the missing energy source in CE evolution. However, quantitative simulations are still needed to accurately determine the contribution of jets to envelope unbinding during the CE phase.}
  
%  The most direct explanation for cases where $\alpha_{\rm CE} > 1$ involves invoking additional energy contributions, such as tidal heating, nuclear burning, or accretion energy. However, as discussed by \citet{Ivanova2013}, these mechanisms likely provide only marginal contributions to envelope ejection. This limitation arises because: (1) the CE phase occurs on extremely short timescales, severely restricting energy transfer efficiency; and (2) the non-degenerate nature of the companion star in these systems inhibits significant supplemental energy generation. Consequently, while these processes may modestly augment the available energy budget, they cannot fully account for the high efficiency values ($\alpha_{\rm CE} \gtrsim 1$) required by our models.

\item \textbf{Tidal effects:} Tidal interactions could indirectly influence the estimated CE efficiency. If sufficiently strong, tides can significantly tighten the orbital separation as the massive star evolves into a RSG. In this scenario, the donor star may not initially fill its Roche lobe at its maximum radial extent, enabling continuous mass loss during the RSG phase as tides gradually shrink the orbit. By the onset of the CE phase (when the star finally fills its Roche lobe), a substantial portion of the envelope may have already been lost. {This process is quite similar to the grazing envelope evolution proposed by \citet{Soker2015}, in which jets remove some envelope mass before the CE phase (see also \citealt{Shiber2024}), thereby substantially reducing the binding energy \citep{Kotko2024}.} However, it is important to emphasize that current tidal theory still contains significant uncertainties, particularly regarding the efficiency of orbital decay \citep[e.g.,][]{Meibom2005,Ogilvie2014,Niej2017}.

\item \textbf{Modified energy mechanism:} In the standard energy formalism adopted in this work, the envelope's energy profile is assumed to remain constant during CE ejection. However, recent studies \citep{Marchant2021,Geh2022,Geh2024} have proposed that the donor star's structure may evolve significantly due to rapid mass loss during the CE phase. In particular, \citet{Geh2022} developed a self-consistent approach to calculate the envelope's binding energy by treating the total energy as a function of the remnant mass. Preliminary results suggest that the CE efficiency need not to be constant, but a function of the initial mass ratio \citep{Geh2024}. Further investigation is required to determine whether this modified framework can consistently explain the formation of BHXBs with $\alpha_{\rm CE} < 1$ while maintaining physical self-consistency.

\item \textbf{Alternative CE description:} Recent work by \citet{Hirai2022} has proposed a novel two-stage CE ejection mechanism for massive binaries, motivated by the stellar structure of RSGs that typically exhibit a substantial radiative layer between their dense He cores and outer convective envelopes. In this model, CE evolution proceeds through distinct phases, i.e., rapid inspiral through the convective envelope, followed by stable mass transfer from the radiative intershell region (see also \citealt{Marchant2021,Gallegos2023,Niey2025}). {Given that the later phase involves stable mass transfer between the core (radiative intershell region) and the companion, the final post-CE separation is determined by the angular momentum loss mechanism and the specific binary parameters prevailing during the mass transfer epoch.} {However, recent simulations by \citet{Cohen2023} suggested that the inner boundary of the envelope convective zone moves deep into the initial radiative zone, highlighting the uncertainties inherent in this formalism. Apart from the energy-based CE prescription, the $\gamma$-mechanism (e.g., \citealt{Nelemans2000,Nelemans2005,Stefano2023}) treats the CE phase via angular momentum conservation. Although the $\gamma$-mechanism appears to reproduce some observed binary systems better than the energy formalism \citep[e.g.,][]{Nelemans2025}, the underlying physical picture remains unclear.} 
%While this framework provides a promising alternative to standard CE prescriptions, its applicability to BHXB formation scenarios requires further verification. %Besides, a totally different CE description based on angular momentum conservation is also introduced in litreatures (e.g., \citealt{Nelemans2000,DiStefano2024}). 

%4. \textbf{CE vs. stable -- enhanced angular momentum loss:} One of the main difficulties in explaning the formation of BHXBs via stable mass transfer processes is that stable MT generally leads to wide separation. 

\item \textbf{Alternative formation channels:} Given {that significant uncertainties remain} in explaining BHXB formation through isolated binary evolution, several studies have proposed alternative pathways, including dynamical interactions in dense stellar environments and three-body systems \citep{Eggleton1986,Naoz2016,Klencki2017,Shariat2025}. These alternative channels are physically viable, but rigorous validation through detailed N-body simulations and comparison with increasingly larger BHXB samples is still required. 
\end{itemize}

%Above all, the BHXB formation puzzles still exist. 
%Above all, fundamental challenges remain in our understanding of BHXB formation mechanisms.
{Clearly, our current understanding of BHXB formation mechanisms remains incomplete, with several fundamental questions still open.}

\section{Conclusions}
\label{sec:5}

This work establishes critical lower limits for CE efficiency parameters ($\alpha_{\rm 0.5U}$, $\alpha_{\rm U}$, $\alpha_{\rm H}$) during massive binary evolution, providing essential benchmarks for population synthesis studies. When accounting solely for internal energy contributions, we find minimum required efficiencies of $\alpha_{\rm 0.5U} \gtrsim 6.7$ and $\alpha_{\rm U} \gtrsim 4.2$ for successful envelope ejection. Notably, including enthalpy, which substantially reduces the envelope binding energy, lowers this threshold to $\alpha_{\rm H} \gtrsim 1.7$, though all values significantly exceed unity, {suggesting that additional energy sources (such as jets) may be necessary. Alternatively, this may indicate that the current description of the CE phase needs to be revised.}

BH natal kicks remain one of the most critical uncertainties in BH formation scenarios. Our analysis reveals that 4U 1543 cannot form through isolated binary evolution with kick velocities below $\sim 50 \;\rm km/s$, with a preferred formation probability peak at $160 \;\rm km/s$. We also systematically investigate how BH natal kicks influence CE efficiency and find that (1) typical $\alpha_{\rm CE}$ values increase with higher kick velocities, (2) all viable solutions maintain $\alpha_{\rm CE} > 1$ regardless of kick magnitude.

A growing consensus suggests that classical energy mechanisms, {which we define here as those relying solely on orbital energy and internal or recombination energy and excluding external sources such as jets,} fail to provide sufficient energy for CE ejection in massive binaries, as supported by the systematic need for $\alpha_{\rm CE}>1$ in population synthesis models attempting to match the observed properties of post-CE massive binaries \citep{Kiel2006,Yungelson2006,Grichener2023}. Our results confirm that while the classical energy formalism offers simplicity and computational convenience, it faces fundamental limitations in describing massive binary CE evolution. These findings highlight the critical need to either modify existing CE prescriptions or develop new theoretical frameworks that properly incorporate the complex physics of massive envelope ejection.

\begin{acknowledgements}
  {We deeply thank the referee for a very careful reading and constructive comments that have led to the improvement of the manuscript. The authors are grateful to Poshak Gandhi for his valuable suggestions and feedback on this work.} This work is supported by the Natural Science Foundation of China (grant Nos. 12125303, 12525304, 12288102, 12473034, 12090040/3, 12273105, 11703081, 11422324, 12073070, 12173081), the CAS Project for Young Scientists in Basic Research (YSBR-148), the Strategic Priority Research Program of the Chinese Academy of Sciences (grant Nos. XDB1160303, XDB1160201, XDB1160000), the National Key R$\&$D Program of China (grant Nos. 2021YFA1600403, 2021YFA1600400), the Yunnan Revitalization Talent Support Program-Science $\&$ Technology Champion Project (No. 202305AB350003) and Young Talent project, the International Centre of Supernovae (ICESUN), Yunnan Key Laboratory of Supernova Research (Nos. 202302AN360001, 202201BC070003), Yunnan Fundamental Research Projects (No. 202401AT070139), the Natural Science Foundation of Henan Province (No. 242300420944). X.C. acknowledges the New Cornerstone Science Foundation through the XPLORER PRIZE. The authors gratefully acknowledge the “PHOENIX Supercomputing Platform” jointly operated by the Binary Population Synthesis Group and the Stellar Astrophysics Group at Yunnan Observatories, Chinese Academy of Sciences. 
%\software{BSE \citep{hurley00,hurley02}}
\end{acknowledgements}
%%%%%%%%%%%%%%%%%%%%%%%%%%%%%%%%%%%%%%%%%%%%%%%%%%
%\begin{contribution}
%%%This section gives authors the space to recognize author contributions. The text inside this environment is NOT counted towards the total word quanta. At a minimum, manuscripts are expected to include this text:
%
%All authors contributed equally to the Terra Mater collaboration.
%
%\end{contribution}

\software{MESA (v12115; \citealt{Paxton2011,Paxton2013,Paxton2015,Paxton2018,Paxton2019,Jermyn2023})}

{Example MESA inlists we used to generate our models are archived on Zenodo and can be downloaded at \dataset[doi:10.5281/zenodo.19660373]{\doi{10.5281/zenodo.19660373}}.}

\newpage

\appendix

\section{The cutoff of the secondary mass}
\label{sec:appA}

\begin{figure}
  \centering
    \begin{minipage}[b]{0.45\textwidth}
        \centering
        \includegraphics[width=\textwidth]{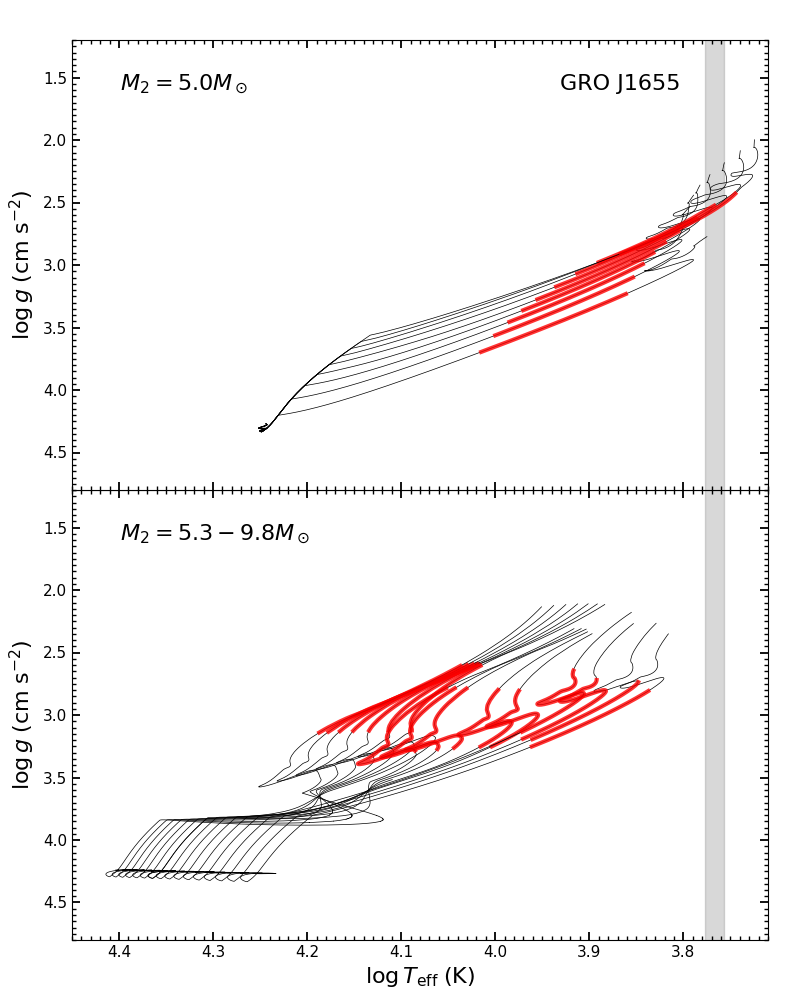}
    \end{minipage}
    \begin{minipage}[b]{0.45\textwidth}
        \centering
        \includegraphics[width=\textwidth]{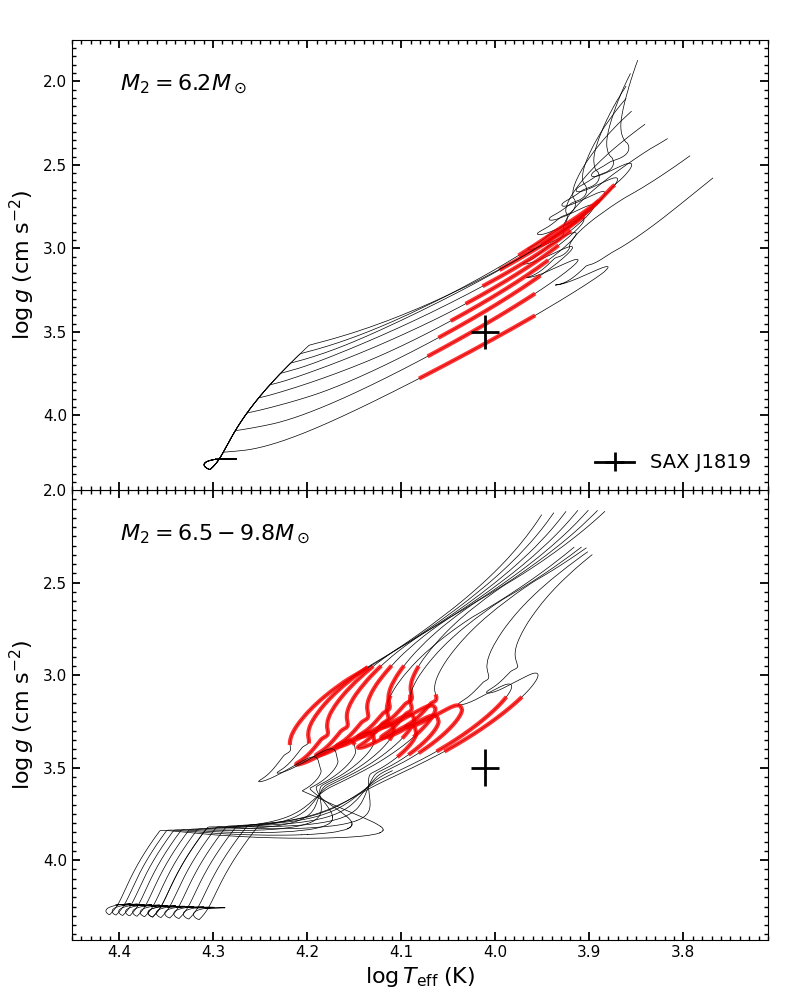}
    \end{minipage}
    \caption{The evolutionary tracks for GRO J1655 (left panels) and SAX J1819 (right panels) as functions of varying secondary masses, which are used to constrain the upper mass limits of the secondary stars. For GRO J1655, we adopt a mass cutoff of 5.0 $M_\odot$, beyond which more massive secondaries cannot reproduce the observed effective temperature of the system. Similarly, for SAX J1819, we implement a cutoff at 6.2 $M_\odot$, as more massive secondaries generally evolve beyond the observed parameter space of this system. The red solid lines indicate cases where the secondary masses fall within the observed ranges for each binary.}
    \label{fig:A1}
\end{figure}

We establish secondary mass cutoffs for GRO J1655 and SAX J1819 based on spectral observations of their companion stars. Figure~\ref{fig:A1} presents the evolutionary tracks for both systems (GRO J1655 in left panels, SAX J1819 in right panels) as functions of secondary mass, which constrain the upper mass limits of these companions. For GRO J1655-40, we impose a 5.0 $M_\odot$ cutoff because more massive companions cannot match the system's observed effective temperature. Similarly, for SAX J1819.8-2523, we adopt a 6.2 $M_\odot$ cutoff since more massive companions exceed the system's observed parameter space.

\section{The CE efficiencies with more massive BH progenitors}
\label{sec:appB}
\begin{figure}
    \centering
    \includegraphics[width=\columnwidth]{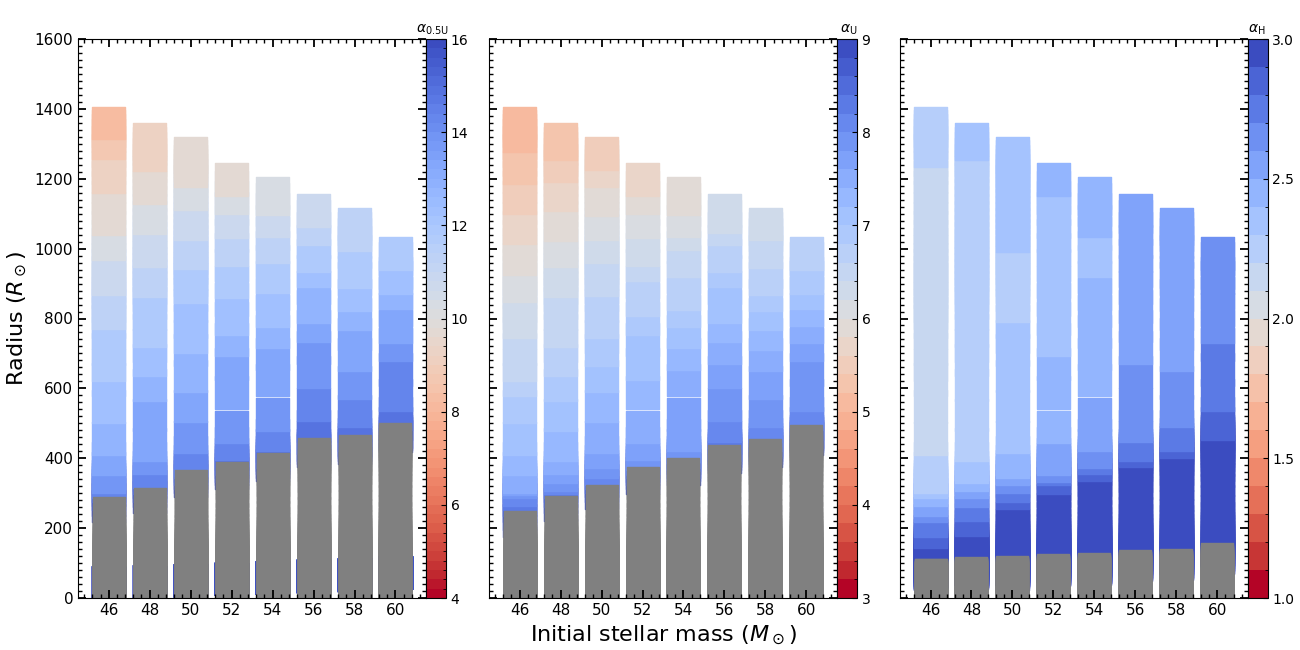}
    \caption{The minimum CE efficiency requirements in the cases of $46-60\,M_\odot$ BH progenitors for GRO J1655. }
    %The sdO first expands towards high luminosity due to the He-shell and H-shell burning.} 
    \label{fig:B1}
\end{figure}

\begin{figure}
    \centering
    \includegraphics[width=\columnwidth]{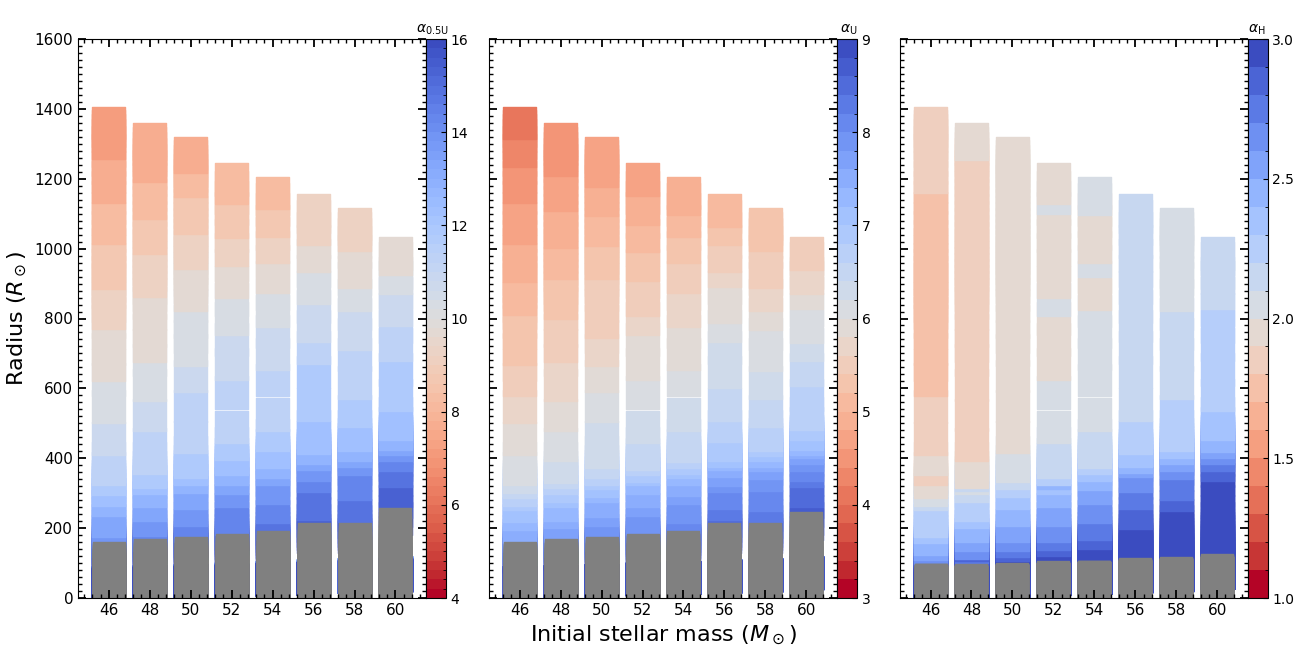}
    \caption{The minimum CE efficiency requirements in the cases of $46-60\,M_\odot$ BH progenitors for SAX J1819. }
    %The sdO first expands towards high luminosity due to the He-shell and H-shell burning.} 
    \label{fig:B2}
\end{figure}

\begin{figure}
    \centering
    \includegraphics[width=\columnwidth]{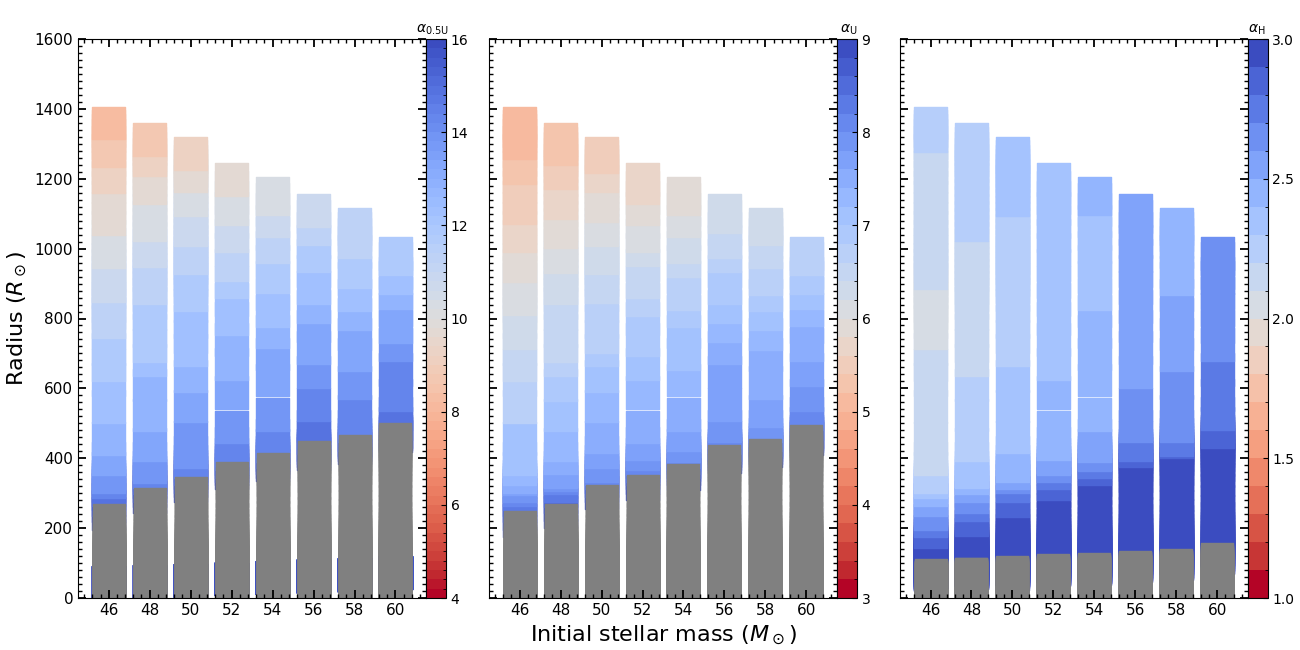}
    \caption{The minimum CE efficiency requirements in the cases of $46-60\,M_\odot$ BH progenitors for 4U 1543. }
    %The sdO first expands towards high luminosity due to the He-shell and H-shell burning.} 
    \label{fig:B3}
\end{figure}

In Figures~\ref{fig:B1}-\ref{fig:B3}, we present the minimum CE efficiency requirements for GRO J1655, SAX J1819, and 4U 1543, respectively, considering more massive BH progenitors. The results suggest systematically higher CE efficiency values compared to systems with lower-mass BH progenitors. 

\section{Minimum CE efficiencies as functions of BH progenitor masses}
\label{sec:appC}

\begin{figure}
    \centering
    \includegraphics[width=\columnwidth]{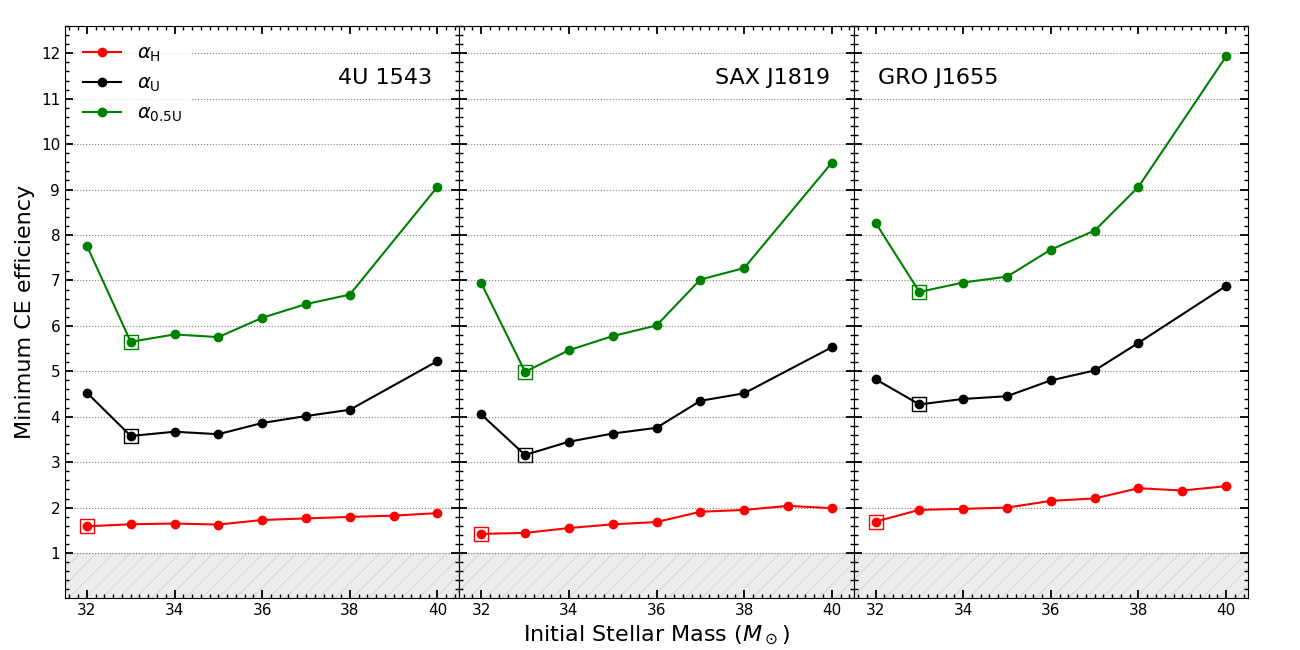}
    \caption{Minimum CE efficiencies as functions of initial BH progenitor masses for the three BHXBs. The open squares denote the corresponding minimum CE efficiency values previously shown in Figure~\ref{fig:13}.}
    %The sdO first expands towards high luminosity due to the He-shell and H-shell burning.} 
    \label{fig:C1}
\end{figure}

Figure~\ref{fig:C1} presents the minimum CE efficiences as functions of initial BH progenitor masses for the three BHXBs. The open squares denote the corresponding minimum CE efficiency values previously shown in Figure~\ref{fig:13}.

\newpage

%% For this sample we use BibTeX plus aasjournalv7.bst to generate the

\bibliography{sample701}{}
\bibliographystyle{aasjournalv7}



\end{document}